\documentclass[10pt,letterpaper]{article}
\usepackage[top=0.85in,left=2.75in,footskip=0.75in]{geometry}

\usepackage{amsmath,amssymb}

\usepackage{changepage}

\usepackage{textcomp,marvosym}

\usepackage{cite}

\usepackage{nameref,hyperref}

\usepackage[right]{lineno}

\usepackage[nopatch=eqnum]{microtype}
\DisableLigatures[f]{encoding = *, family = * }

\usepackage[table]{xcolor}

\usepackage{array}

\newcolumntype{+}{!{\vrule width 2pt}}

\newlength\savedwidth

\raggedright
\usepackage[aboveskip=1pt,labelfont=bf,labelsep=period,justification=raggedright,singlelinecheck=off]{caption}

\makeatletter
\renewcommand{\@biblabel}[1]{\quad#1.}
\makeatother

\usepackage{lastpage,fancyhdr,graphicx}
\usepackage{epstopdf}
\fancyheadoffset[L]{2.25in}
\fancyfootoffset[L]{2.25in}
\usepackage{graphicx} 
\usepackage{dcolumn} 
\usepackage{bm} 

\usepackage{hyperref} 

\usepackage[linesnumbered, ruled, vlined]{algorithm2e}

\usepackage{xcolor}

\usepackage{amsthm}

\usepackage{xr}

\usepackage{slashed}

\usepackage{tikz}

\usepackage{enumitem}

\usetikzlibrary{quantikz}

\usepackage{xcolor}

\allowdisplaybreaks

\usepackage{MnSymbol}

\usepackage{soul}

\makeatletter
\newcommand{\vast}{\bBigg@{3.0}}
\newcommand{\Vast}{\bBigg@{3.5}}

\begin{document}
\vspace*{0.2in}

\begin{flushleft}
{\Large
\textbf\newline{Emergent invariance and scaling properties in the collective return dynamics of a stock market} 
}
\newline
\\
Hideyuki Miyahara\textsuperscript{1},
Hai Qian\textsuperscript{1},
Pavan Holur\textsuperscript{1},
Vwani Roychowdhury\textsuperscript{1*}
\\
\bigskip
\textbf{1} Department of Electrical and Computer Engineering, University of California, Los Angeles, CA, 90095, USA
\bigskip

%
%





* vwani@ucla.edu

\end{flushleft}
\section*{Abstract}

A key metric to determine the performance of a stock in a market is its \textit{return} over different investment horizons ($\tau$). Several works have observed heavy-tailed behavior in the distributions of returns in different markets, which are observable indicators of underlying complex dynamics. Such prior works study return distributions that are marginalized across the individual stocks in the market, and do not track statistics about the joint distributions of returns conditioned on different stocks, which would be useful for optimizing inter-stock asset allocation strategies. As a step towards this goal,  we study emergent phenomena in the distributions of returns as captured by their pairwise correlations. In particular, we consider the pairwise (between stocks $i,j$) partial correlations of returns with respect to the market mode, $c_{i,j}(\tau)$, (thus, correcting for the baseline return behavior of the market), over different time horizons ($\tau$),  and discover two  novel emergent phenomena: (i) the standardized distributions of the $c_{i,j}(\tau)$'s are observed to be invariant of $\tau$ ranging from from $1000 \textrm{min}$ (2.5 days) to $30000 \textrm{min}$ (2.5 months); (ii) the scaling of the standard deviation of $c_{i,j}(\tau)$'s with $\tau$ admits good fits to simple model classes such as a power-law $\tau^{-\lambda}$ or stretched exponential function $e^{-\tau^\beta}$ ($\lambda,\beta > 0$). Moreover, the parameters governing these fits provide a summary view of market health: for instance, in years marked by unprecedented financial crises -- for example $2008$ and $2020$ -- values of $\lambda$ (scaling exponent) are substantially lower. Finally, we demonstrate that the observed emergent behavior cannot be adequately supported by existing generative frameworks such as single- and multi-factor models. We introduce a promising agent-based Vicsek model that closes this gap.

\section*{Background}
Stock prices demonstrate considerable volatility, a result of several confounding factors such as traders’ collaborative and competitive decision-making to buy, hold or sell, differing appetites for risk, and various time horizons for expected returns on investment 
~\cite{Newman_001, Ladyman_001, Ladyman_002, Mantegna_001, Bonanno_001, Bouchaud_001, Stanley_001}. A considerable body of literature has focused on identifying patterns in price fluctuations and on developing succinct dynamical models that display similar characteristics as a real market. Financial experts have, for instance, frequently observed seasonality patterns in individual stock prices and the fractal nature of price fluctuations~\cite{seasonality, mandelbrot}. 
In contrast, macroscopic patterns that concern the ensemble of stocks 
would emerge because of correlated dynamics in investment decisions, and would reflect   inter-stock asset allocation strategies used by investors. The  abundant literature in swarming and ``econophysics''~\cite{swarmed, Mantegna_001} provide a framework for both numerical analysis of such joint price data and for formulating theoretical generative models.

Conventionally, investors and economists compute the \textit{return}~\cite{reviewer_1}: the return over an investment horizon of $\tau$ is defined as the equivalent compounded interest rate if one bought a stock at time $t$ and sold it at time $t+\tau$, and estimated as $r(t;\tau) := \ln(p(t+\tau)) - \ln(p(t)) $, where $p(t)$ and $p(t+\tau)$ are the stock prices at time $t$ and $t+\tau$, respectively. 
Return sequences are a stationary measure (a parameter constant over the interval $\tau$) of price change characterized by several statistical properties observed in empirical evidence across markets (often referred to as \textit{stylized facts}~\cite{wpenn}).

Prior research has explored the statistical properties of these returns as a means to characterize the dynamics of the market. Some~\cite{reviewer_1, drozdz} have identified linear relationships in the log-log scale on the distributions of the returns. Plerou et al.~\cite{plerou_004} discover power law fits~\cite{reinventwillis, experience} on (i) the cumulative distribution function (CDF) of the return distribution and (ii) the standard deviation of the return distribution as a function of market capitalization. Similarly, Müller et al.~\cite{muller} demonstrated evidence of a scaling regime governing the mean of returns with respect to the return horizon (for $\tau \le 20$ seconds). Return distributions and their properties for longer horizon $\tau$ beyond high frequency trading time scales have not been studied. \textit{Moreover, such prior works only study return distributions that are marginalized across the individual stocks in the market}. For example,  in \cite{reviewer_1, drozdz}, they fix a $\tau$ and then compute $r_i(\tau)$ for all stocks $i$ in a market and then estimate the distribution of this set of $r(\tau)$'s; thus marginalizing over all the stocks, and the Complementary CDFs (CCDFs) for our dataset is presented in Fig.~\ref{fig:ccdf} of the supplementary material and for large enough $\tau$ the tails can be fit with that of power-law distributions. Similarly, in \cite{plerou_004}  they compute the standard deviation, $\sigma_i(\tau)$, of returns for a fixed $\tau$ and for all stocks with market capitalization $S_i$ and then they show that $\log(\sigma_i(\tau)$ scales linearly with $\log(S_i)$; again, the returns are marginalized over all the stocks with the same capitalization. \textit{None of these works 
track statistics about the joint distributions of returns} conditioned on different stocks, which would be useful for optimizing inter-stock asset allocation strategies. 

To address such issues, other methods have attempted to model the inter-stock return correlations in order to compare stocks' relative performances over time \cite{Mantegna_001}. These correlations are particularly useful to construct \textit{graphical models} of the market, in this case, a fully connected network, where the stocks are nodes and the pair-wise correlations correspond to the inter-stock edge weights. Structures are distilled from within the network representation by adopting various graph theory algorithms \cite{Tumminello_001, Aste_001, Tumminello_002, Onnela_001}. For example, one can compute the Minimum Spanning Tree (MST), which, for certain return horizons, exhibit a local aggregation of communities of stocks (nodes), such that each community is shared by stocks belonging to market sector~\cite{reviewer_1, kwapien}. Recent studies have attempted to refine this method of identifying clusters  by calculating the normalized \textit{partial} correlations in relation to the market mode~\cite{korean}. With \textit{fixed} $\tau > \tau_0$ -- where the MST structures are obtained -- these partial correlation scores have been observed to contain enough inter-stock information to facilitate agglomerative clustering of the Korean Stock Exchange (KOSPI) that aligned with GICS sectors, while network modeling of the partial correlations computed using daily returns have helped uncover specific stocks that are influential in driving the return patterns in a subset of high-capitalization NYSE stocks~\cite{plosone}.

\section*{Our Contributions} \label{contributions}

Such MST-based studies (see Fig.~\ref{main_numerical_result_ensemble_00001_years_00001_mst_001_001}) provide only a limited visual representation of the underlying return correlation distribution and its dependence on $\tau$. We extend such MST-based analysis of the correlation statistics to the study of its density function, governing the return correlations. In contrast to existing work discussed earlier, we are interested in: (a) the distributions of \textit{market-mode adjusted} partial correlations computed at both short ($1$ minute) and \textit{long} investment horizons ($50000$ minutes); and (b) the distribution properties as a function of the investment horizon $\tau$. \textit{ As our first main contribution}, we find that  for a significant range of $\tau$ (varying from approximately  2.5 days to around 2.5 months), the standardized distributions -- scaled by the standard deviation $\sigma(\tau)$ and zero-shifted by mean $\mu(\tau)$-- of the market-mode adjusted partial return correlations are invariant of $\tau$.

The above distribution invariance results suggest that both the standard deviation $\sigma(\tau)$ and the mean $\mu(\tau)$ of the pairwise partial correlations would be a function of return horizon $\tau$ in the regime where such invariance is observed. We find that $\mu(\tau)$ has no significant scaling behavior with respect to $\tau$ (see Fig.~\ref{fig:means} in the Supplementary Material).   \textit{As our second  main contribution}, we find that  the standard deviations of the partial return correlations do indeed scale as a function of $\tau$ in the distribution invariance regime, and demonstrate convincing fits via either a power law or a stretched exponential function. The critical model parameters -- the scaling exponent in the case of the power law ($\lambda$), and the stretching parameter ($\beta$) in the stretched exponential function -- appear to be rich indicators of macroeconomic volatility patterns. The distribution invariance as well as the scaling of $\sigma(\tau)$ are observed to hold for $1000 \textrm{min} \leq \tau \leq 30000 \textrm{min}$. Evidence spans $17$ years of real S\&P500 stock price data, sampled every minute. Data can be accessed for research purpose at the Wharton Research Data Services (WRDS)\footnote{\url{https://wrds-www.wharton.upenn.edu/}} and the code repository \footnote{\url{https://github.com/pholur/stock-market}} is linked.


Finally, we explore if these numerically observed emergent behavior properties can be replicated by an agent-based generative model. We first reexamine the single- and multi-factor generative models, popular generative frameworks used to model consensus behavior in financial markets~\cite{baselinefactor}. These models for the most part fail to replicate the above-mentioned emergent trends -- the invariant standardized histograms and the power-law/stretched-exponential fits with respect to $\tau$: (a) The single-factor model fails to reproduce the vine MST structure; (b) Multi-factor models, while generating the vine, fail to produce both the distribution-invariance and the scaling phenomena. \textit{As our third main contribution,} we introduce an alternate framework, a modified Vicsek model -- commonly used to describe the dynamics of active matter -- that proves to be a much more promising candidate for reproducing the empirical evidence. In these approaches, the stock market would be modeled as a closed environment, where individual stocks behave as agents in a vector space that influence each other. At each time-step, the position of an agent corresponds to a particular stock's instantaneous market behavior. Agents that exhibit correlated behavior over multiple time steps cluster together as swarms.


\section*{Materials and Methods}

\subsection*{Correlations of returns and partial correlation with respect to market mode} \label{sec:correlation}

Suppose a market has $N$ stocks; in the S\&P500, $N \approx 500$. Let us denote, by $p_i (t)$, the price of stock $i$ at time $t$ for $i = 1, 2, \dots, N$ and $t_\mathrm{ini} \le t \le t_\mathrm{fin}$. Typical macroeconomic market analyses such as Year-over-Year (YoY) gain, annualized returns and GDP growth, cap $T_{int} \coloneqq t_\mathrm{fin} - t_\mathrm{ini}$ to $1$ year from January 1st to December 31 to avoid seasonality patterns and resulting artifacts in  the correlations. We similarly consider each calendar year separately and the evidence of scaling is thus presented individually for each of the $17$ years ($2004-2020$).

We sample each of the stock prices at a granularity of $1-$minute. Let the price sequence of a stock $i$ be $p_i$. For $T_{int} = 1 \, \textrm{year}$, there are $\sim 98000$ values per sequence. We compute the effective return or interest rate $r_i(t;\tau)$ of the stock $i$ at time $t$ over a time horizon of $\tau$ ($\tau \le T_{int}$), a preferred \textit{first-order} metric for investing than the absolute price. An investment in the $i^{th}$ stock at time $t$ (say, a sum of $mp_i(t)$  by purchasing $m$ units) when \textit{compounded continuously} at the given rate  would yield the same amount as that which would be obtained by selling the stock at time $(t+\tau)$ (i.e., $m p_i(t+\tau)$). Quantitatively, $\displaystyle mp_i(t+\tau) = m p_i(t) \lim_{n \rightarrow \infty} \left( 1+\frac{r_i(t;\tau)}{n\tau}\right)^{n\tau} = m p_i(t) e^{r_i(t;\tau)}$. Thus, we get:
\begin{align}
r_i (t; \tau) &\coloneqq \ln p_i (t + \tau) - \ln p_i (t),
\label{main_eq_def_interest_rate_001_001}
\end{align}
for $t_\mathrm{ini} \le t \le t_\mathrm{fin} - \tau$. Therefore, an investment horizon of $\tau$ yields a return sequence of length, $T_{int} - \tau + 1$. Note that for the longest considered $\tau = 30000 \, min$, the return sequence for each stock still contains a significant number of return values ($\sim 68000$). After computing the return sequences of all stocks, we can find the market-mode return sequence as:

\begin{align}
  r_0 (t; \tau) &\coloneqq \frac{1}{N} \sum_{i=1}^N r_i (t; \tau).
\end{align}
We denote the time average of $r_i (\cdot; \tau)$ and $r_i (\cdot; \tau) r_j (\cdot; \tau)$ for $i, j = 0, 1, 2, \dots, N$ by
\begin{align}
\overline{r_i (\cdot; \tau)} &\coloneqq \int_{t_\mathrm{ini}}^{t_\mathrm{fin} - \tau} dt \, r_i (t; \tau), \\
\overline{r_i (\cdot; \tau) r_j (\cdot; \tau)} &\coloneqq \int_{t_\mathrm{ini}}^{t_\mathrm{fin} - \tau} dt \, r_i (t; \tau) r_j (t; \tau). \label{main_def_correlation_ij_001_001}
\end{align}
Now we are ready to define the \textit{conventional correlation}. Between any pair of stocks $i, j = 1, 2, \dots, N$:
\begin{align}
\rho_{i, j} (\tau) &\coloneqq \frac{\sigma_{i, j} (\tau)}{\sigma_i (\tau) \sigma_j (\tau)}, \label{main_def_conventional_correlation_001_001}
\end{align}
where
\begin{align}
\sigma_i (\tau) &\coloneqq \sqrt{\overline{r_i^2 (\cdot; \tau)} - \overline{r_i (\cdot; \tau)}^2}, \\
\sigma_{i, j} (\tau) &\coloneqq \overline{r_i (\cdot; \tau) r_j (\cdot; \tau)} - \overline{r_i (\cdot; \tau)} \cdot \overline{r_j (\cdot; \tau)}.
\end{align}

\noindent Note: In an optimized market, the cross-correlation between $r_i (t)$ and $r_j (t + \Delta)$ for non-zero $\Delta$ can be written by replacing the right-hand side of Eq.~\eqref{main_def_correlation_ij_001_001} by:
\begin{align}
  \int_{t_\mathrm{ini}}^{t_\mathrm{fin} - \tau - \Delta} dt \, r_i (t; \tau) r_j (t + \Delta; \tau).
\end{align}
However, these correlation should equal $0$; otherwise investors would use one return series to predict another stock’s return (recall Stylized Fact I~\cite{wpenn}).


Next we introduce the concept of partial correlation between stocks $i$ and $j$ with respect to the market mode~\cite{PC}: let
$\tilde{r}_i(t;\tau)$ be the residuals while predicting $r_i(t;\tau)$ with respect to 
$r_0(t;\tau)$ using a linear fit. Then the correlation between these residuals associated to stocks $i$ and $j$ is the partial correlation and is given by,

\begin{align}
  c_{i, j} (\tau) &\coloneqq \frac{\rho_{i, j} (\tau) - \rho_{i, 0} (\tau) \rho_{j, 0} (\tau)}{\sqrt{1-\rho_{i,0}^2}\sqrt{1-\rho_{j,0}^2}}. \label{main_def_nonmarket_correlation_001_001}
\end{align}

where $\rho_{i,j}(\tau)$ is the (conventional) correlation between stocks $i$ and $j$, and $\rho_{i, 0}$, $\rho_{j, 0}$ are correlations of returns of stocks $i$ and $j$ with respect to the market return.


\subsection*{Distributions and invariance of partial correlations}

\begin{figure}[!h]
\centering
\includegraphics[width=0.70\columnwidth]{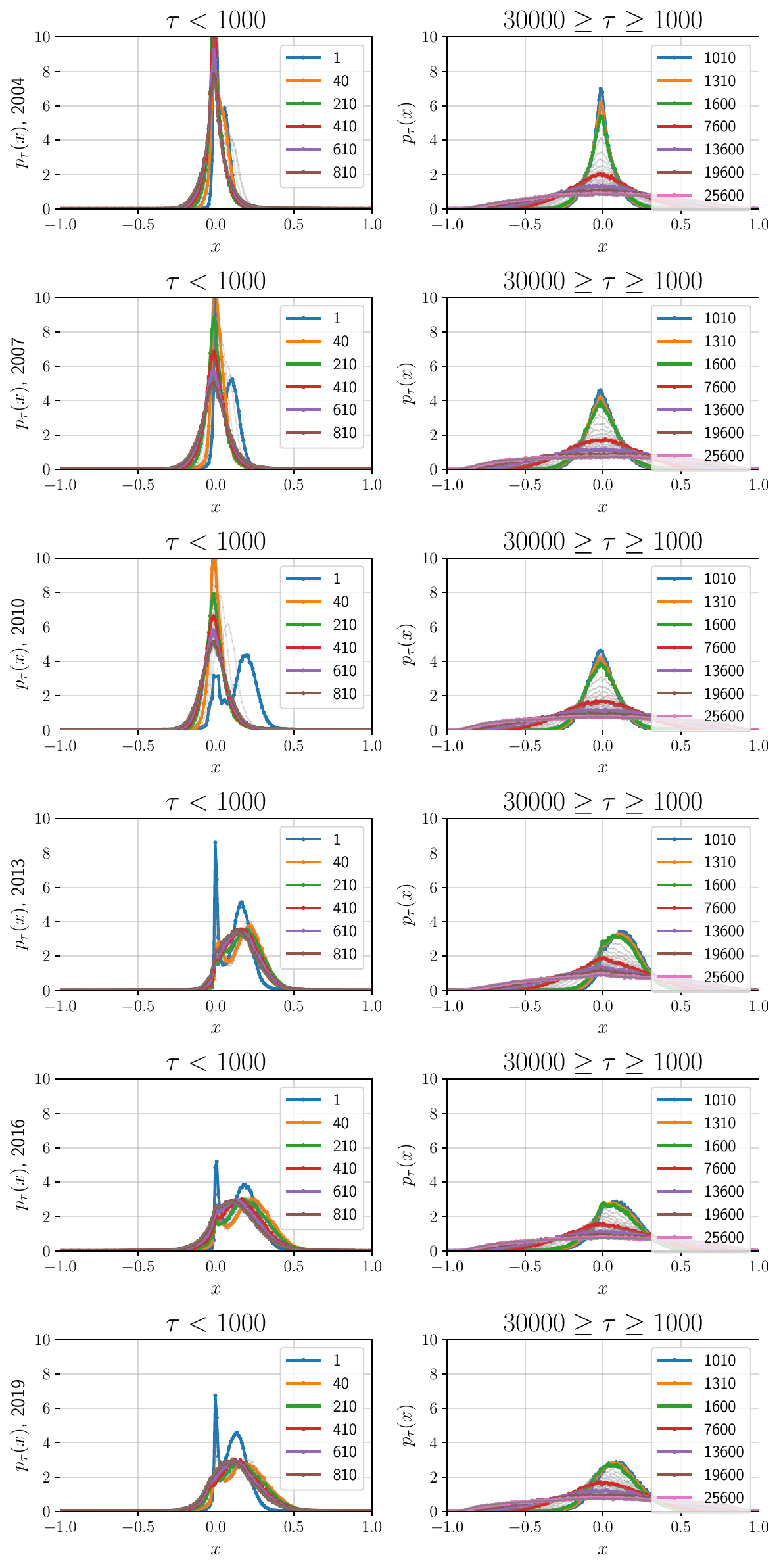}
\caption{{\bf Empirically-estimated probability density function $p_\tau(x)$ of $c_{i,j}(\tau)$s:} The PDF defined in Eq.~\eqref{main_def_p_tau_c_001_001} is visualized for $6$ years - 1 year per row: (a) Left: $\tau = \text{1 min}$ to $\text{1000 min}$ and (b) Right: $\tau = \text{1000 min}$ to $\text{30000 min}$. For $\tau > \tau_0$, the normalized PDFs are invariant to $\tau$ (see Fig.~\ref{main_numerical_result_ensemble_00001_years_00001_cij_scaled_raw_data_001_001, main_numerical_result_ensemble_00001_years_00001_cij_scaled_raw_data_001_002}).}
\label{fig:raw_histograms}
\end{figure}

Let $p_\tau(x)$ be the probability density function of $c_{i,j}(\tau)$ empirically estimated as:

\begin{align}
  p_\tau (x) \sim \frac{2}{N(N-1)} \sum_{i < j} \delta_\mathrm{D} (x - c_{i, j} (\tau)), \label{main_def_p_tau_c_001_001}
\end{align}
where $\delta_\mathrm{D} (\cdot)$ is the Dirac delta function. We observe that as to be expected, the functional form of $p_\tau (\cdot)$ is $\tau$-dependent (see Fig.~\ref{fig:raw_histograms}). However,   \textbf{the standardized distributions} -- the distribution when $c_{i,j}(\tau)$ are mean-shifted and scaled by standard deviation,
\begin{align}
  \tilde{p}_\tau (x) &\coloneqq \frac{1}{b (\tau)} p_\tau \bigg( \frac{x - m(\tau)}{b (\tau)} \bigg), \label{main_def_scaled_distribution_cij_001_001}
\end{align}
\textbf{are invariant over a significant regime of $\tau$}; i.e. $\tau_{\textrm{max}} \ge \tau \ge \tau_{\textrm{min}}$. This indicates that the scaling factor -- in this case, the standard deviation -- scales with $\tau$.

\subsection*{Scaling phenomena during distribution invariance}

Let $\sigma(\tau)$ be the standard deviation and $m(\tau)$ the mean. In this work, we use the \textit{inverse} of the standard deviation -- rather than the standard deviation -- as an interpretable measure of \textit{precision}, $b(\tau) = \frac{1}{\sigma(\tau)}$.

We will demonstrate that during the regime where the \textit{standardized} distributions are invariant ($\tau > \tau_0$), the dependence between $\tau$ and the precision $b(\tau)$ is well-explained by simple models with few and interpretable parameters such as the power law,

\begin{align}
  b (\tau) \sim \tau^{-\lambda}, \quad \textrm{or} \quad \log b(\tau) \approx -\lambda \log(\tau) + C \label{main_eq_scaling_b_001_001}
\end{align}
and the stretched exponential function:
\begin{align}
  b (\tau) \sim e^{\tau^{-\beta}}, \quad \textrm{or} \quad \log b(\tau) \approx \alpha \tau^{-\beta} + \gamma \label{main_eq_scaling_b_001_002}
\end{align}

Aside from convincing model fits, the critical model parameters -- the scaling exponent $\lambda$ (in the case of the power law) and the stretching parameter $\beta$ (in the case of the stretched exponential) -- once trained, emerge as candidates for macroeconomic indicators of market volatility. Indeed, we suspect that \textit{any other simple model class that can convincingly fit and validate the (near-linear in log-log scale) dependence of $b(\tau)$ with respect to $\tau$ should similarly express the market characteristics within its model parameters.}

\section*{Empirical Results: Emergent phenomena 
 in real-world data}

Recall the dataset descriptors: $T_\mathrm{int} = 1$ year; the sampling time interval of stock prices is $1$ minute; there are $\sim251$ business days in a year when the stock market is open from $9:30$ to $16:00$ ET; for every day the market is open, each stock has $\sim390$ prices. For $T_\mathrm{int} = 1$ year, each stock has a $\sim98000$-length price series; the price series is arranged such that the closing price at $16:00$ ET on the current market day immediately precedes the opening price at $9:30$ ET on the following market day. While volatility in after-hours trading may result in drastic price fluctuations at particular indices in the series, an increasing $\tau$ has a smoothing effect on these spikes, and we believe that the return correlation PDFs are not significantly affected by these gaps. Results presented below are replicated for a shorter $T_{int} = 3$ months (see \nameref{S1_Exp_3}).


\subsection*{Functional form of the standardized partial correlation PDF is invariant during a finite $\tau$ regime}
\label{functional}

We provide qualitative and quantitative evidence of the invariance of the functional form of the standardized distribution for a finite regime $30000\textrm{min} > \tau > 1000\textrm{min}$. First, in Fig.~\ref{main_numerical_result_ensemble_00001_years_00001_cij_scaled_raw_data_001_001, main_numerical_result_ensemble_00001_years_00001_cij_scaled_raw_data_001_002}, we plot the standardized histograms for $6$ years (remaining years can be verified using the attached codebase), by superimposing the functions $\tilde{p}_\tau (\cdot)$ across different $\tau$. Quantitative evidence is provided next:

\begin{figure}[!h]
\centering

\includegraphics[width=0.70\columnwidth]{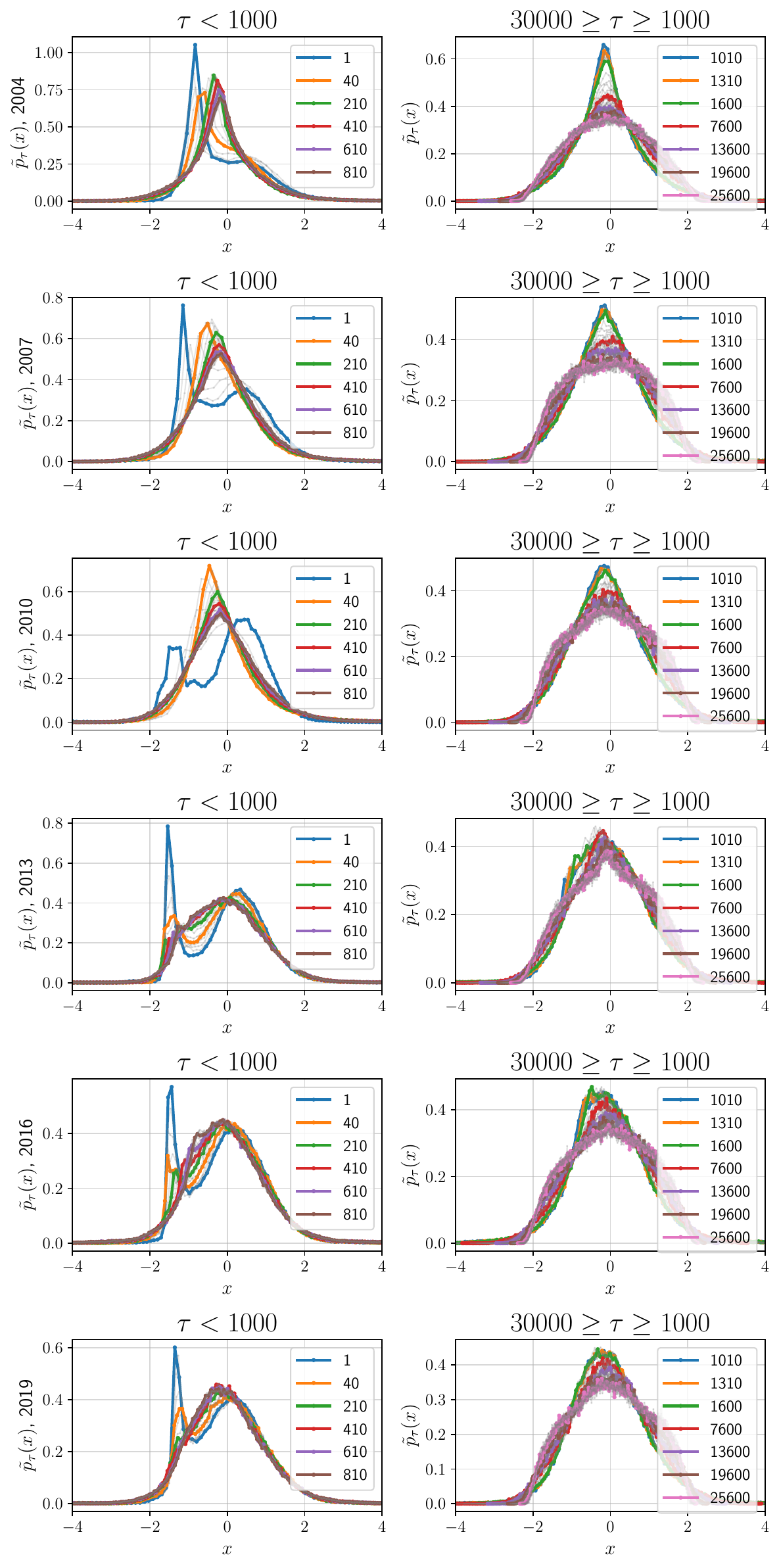}
\caption{{\bf Qualitatively demonstrating the stability of the functional form for increasing $\tau$}: Standardized PDF $\tilde{p}_\tau (\cdot)$ in Eq.~\eqref{main_def_scaled_distribution_cij_001_001} visualized for $6$ years - 1 year per row: (a) Left: $\tau = \text{1 min}$ to $\text{1000 min}$ and (b) Right: $\tau = \text{1000 min}$ to $\text{30000 min}$. As $\tau$ exceeds $\text{1000 min}$, the shape of $\tilde{p}_\tau (\cdot)$ takes a more stable form. A similar analysis with $T_{int} = 3$ months is presented in the Appendix Fig.~\ref{supp:functional_form}.}
\label{main_numerical_result_ensemble_00001_years_00001_cij_scaled_raw_data_001_001, main_numerical_result_ensemble_00001_years_00001_cij_scaled_raw_data_001_002}
\end{figure}

\begin{itemize}
\item \textbf{Pairwise KL divergence between standardized partial correlation PDFs}: \label{pkl}
For each year from $2004$ to $2020$, we compute the KL divergence (KLD) between $\tilde{p}_{\tau_1}(\cdot)$ and $\tilde{p}_{\tau_2}(\cdot)$, the standardized partial correlation PDFs computed with $\tau_1$ and $\tau_2$ respectively. We would like to show that inside the regime where functional invariance was visually observed ($1000\textrm{min} \leq \tau \leq 30000\textrm{min}$), $D_{KL}(\tilde{p}_{\tau_1} || \tilde{p}_{\tau_2})$ for any pair $\{\tau_1, \tau_2\}$ is small compared to the KLD computed between a pair of standardized PDFs for which $\tau$ is outside the scaling region. The pairwise KL divergence between the standardized partial correlation PDFs across $6$ evaluated years are presented in Fig.~\ref{fig:dis-kl_div}. The dark square block in the bottom right of each heatmap implies that the KL divergence between any pair of standardized distributions sampled from the region of $\tau$ specifying the functional invariance is low. In order to compute the KL divergence in a consistent and comparable fashion, each standardized PDF is re-sampled ($N = 10000$) using Gaussian smoothing, $\mathcal{N}(0, 0.2)$.

\begin{figure}[t]
\centering
\includegraphics[width=1.0\columnwidth]{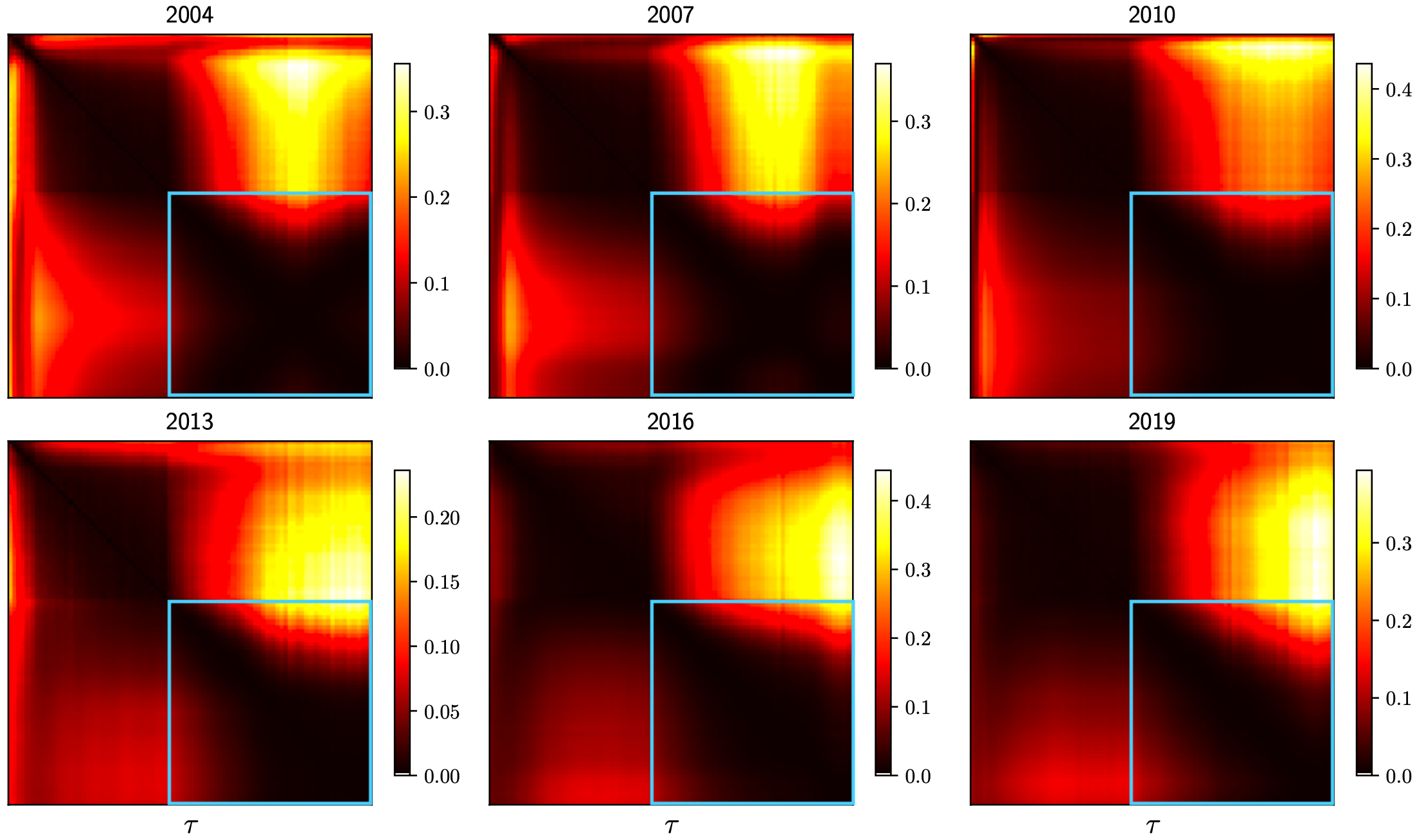}
\caption{{\bf Pairwise KL divergence between standardized partial correlation PDFs for $6$ years of analysis:} Dark blocks along the major diagonal (circumscribed in blue) indicate that when the invariance is visually observed ($1000 \leq \tau \leq 30000$), the pairwise KL divergence is low.}
\label{fig:dis-kl_div}
\end{figure}

\begin{figure}[!h]
\centering
\includegraphics[width=1.0\columnwidth]{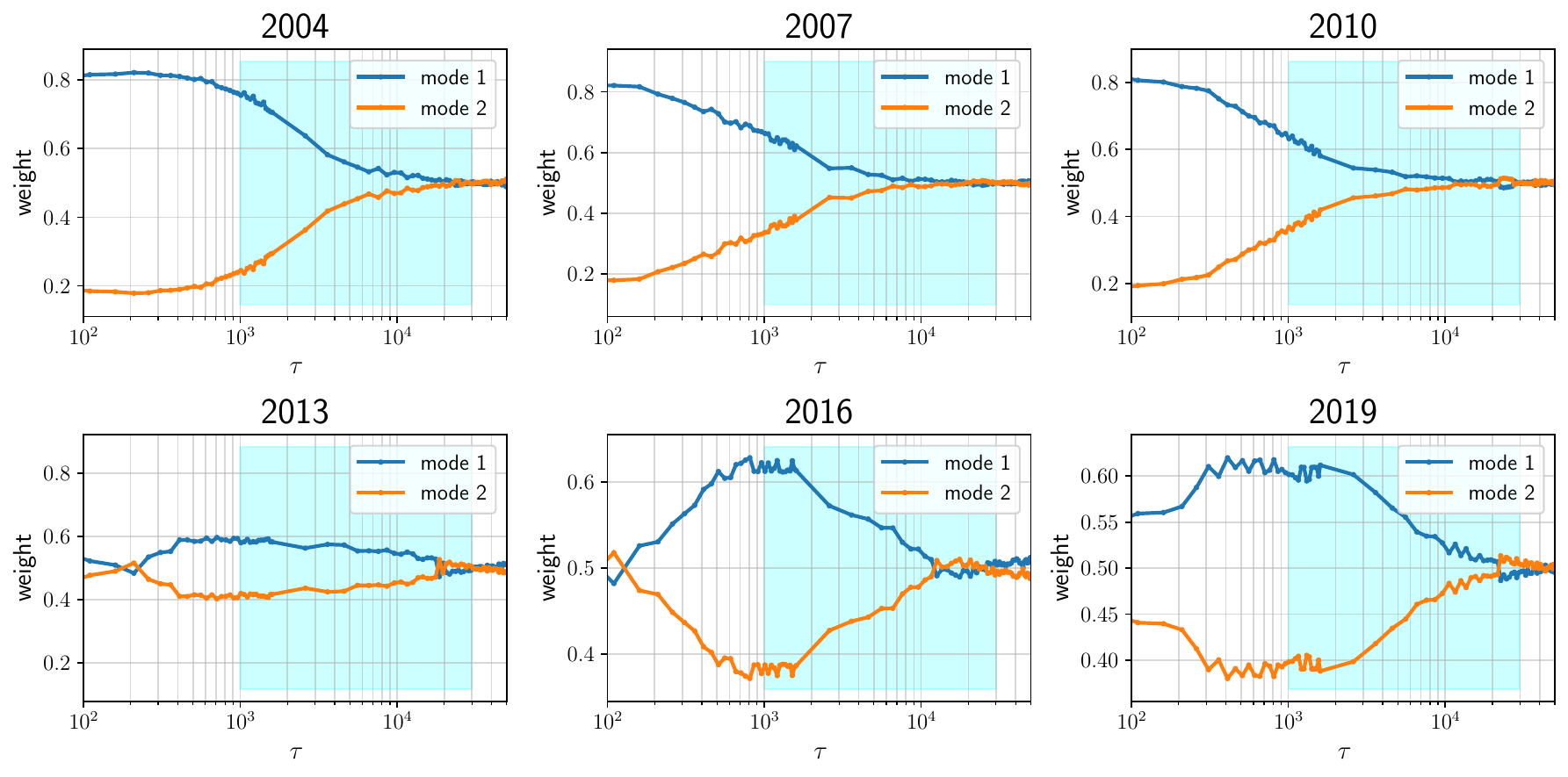}
\caption{{\bf Capturing the onset of function invariance  by fitting 2-mode GMM onto the standardized partial correlation PDFs:} Across $6$ years of analysis, we plot the weights of the $2$ GMM components after fitting to $\tilde{p}_\tau(\cdot)$ for different $\tau$. Mode $1$ corresponds to the mode with the lower standard deviation. As $\tau$ is increases, the second mode starts contributing significantly to the fit signaling the onset (shaded cyan).}
\label{fig:dis-weights}
\end{figure}

\item \textbf{Probing the onset of the function invariance using Gaussian Mixture Models and Kurtosis}: We fit a Gaussian Mixture Model (GMM) ($2$-mode) on $\tilde{p}_\tau(\cdot)$ and probe the weights of the two components across $\tau$. We expect to see a transition as the function invariance sets in. As shown in Fig.~\ref{fig:dis-weights}, initially, one mode is dominant, and as $\tau > 1000$, the weights of the two modes become comparable. Such transitions are observed with $3,4,5$-mode fits as well.

\item \textbf{Kurtosis of the density function with respect to $\tau$}: In Fig.~\ref{fig:dis-kurt}, we demonstrate the same transition (from $\tau < 1000$min to $\tau > 1000$min) by plotting the kurtosis of $\tilde{p}_\tau(\cdot)$ with respect to $\tau$. When the invariance property takes effect, the kurtosis values suggest a corresponding transition from a leptokurtic ($>3$) to platykurtic ($<3$) regime.

\begin{figure}[!h]
\centering
\includegraphics[width=0.8\columnwidth]{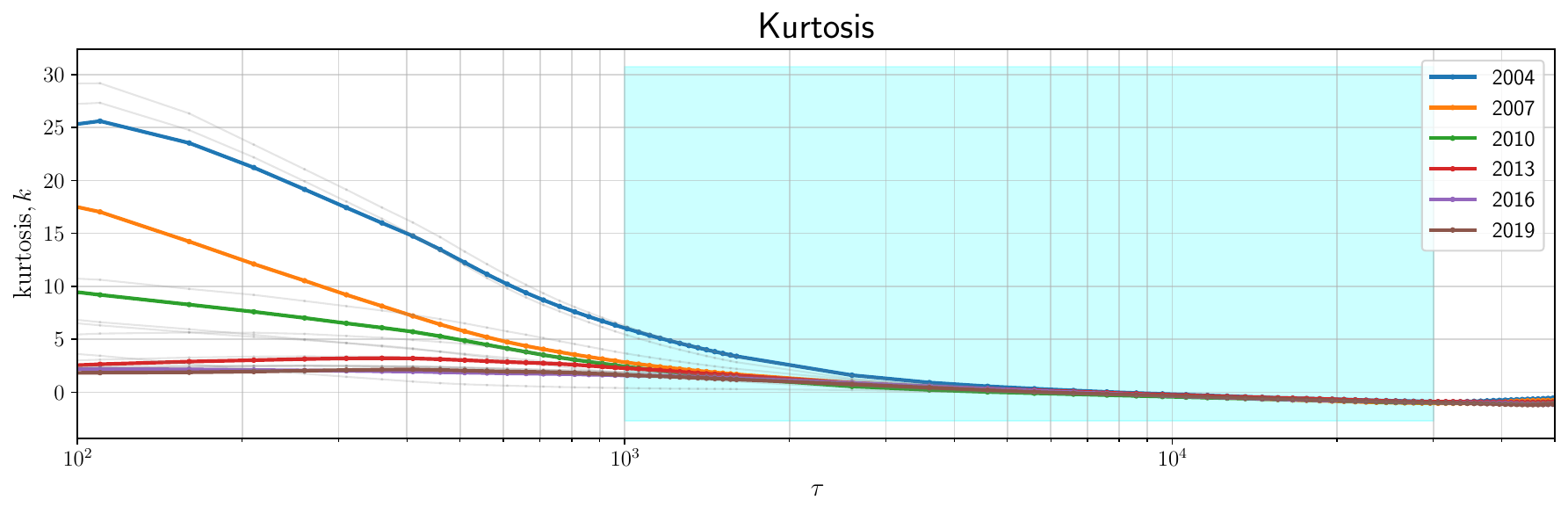}
\caption{{\bf Leptokurtic-to-platykurtic kurtoses transition in the correlation PDFs as the scaling (shaded cyan) takes effect:} The higher values of kurtosis for smaller $\tau$ indicate sharper peaks in the correlation PDFs (compared to a normal distribution) (leptokurtic regime). As $\tau$ increases, the correlation distribution becomes more flat resulting in lower kurtosis (platykurtic regime).}
\label{fig:dis-kurt}
\end{figure}

\end{itemize}

\subsection*{The scaling behavior and its emergent properties}
\label{scaling_law_section}

\begin{figure}[h]
\centering
\includegraphics[width=1.0\columnwidth]{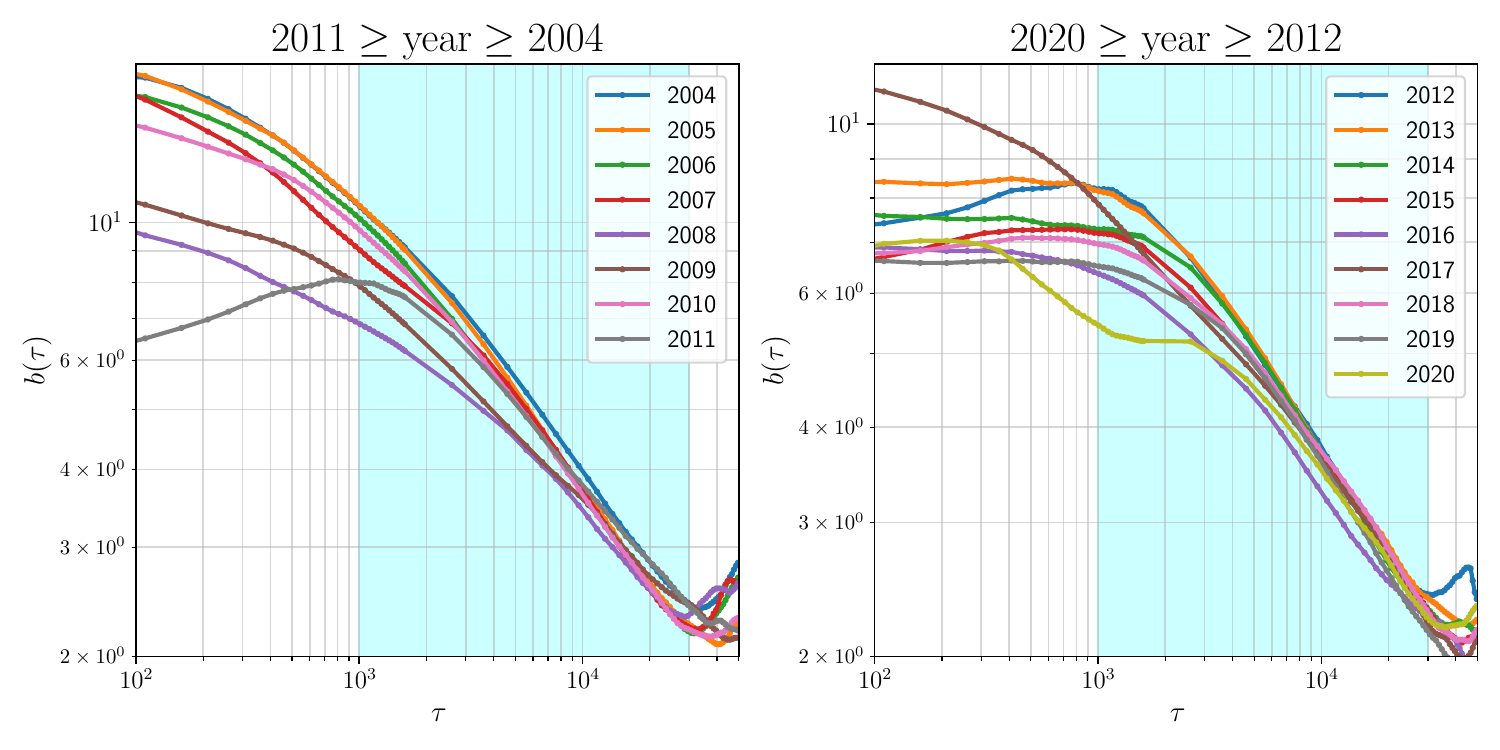}
\caption{{\bf Dependence of the scaling factor $b (\tau)$ on $\tau$:} (a) from 2004 to 2011; and (b) from 2012 to 2020. From $\tau = \text{1000 min}$ to $\text{30000 min}$ (the regime highlighted by light cyan), observe the near-linear relationship between $\ln \tau$ and $\ln b$. A similar visualization with $T_{int} = 3$ months is presented in the Appendix Fig.~\ref{supp:scaling_law}.}
\label{main_numerical_result_ensemble_00001_years_00001_scaling_law_001_001}
\end{figure}

Motivated by the observed function invariance in the standardized distributions of the partial return correlations, in Fig.~\ref{main_numerical_result_ensemble_00001_years_00001_scaling_law_001_001} we plot $b(\tau)$ -- the precision (defined in Materials and Methods) -- as a function of $\tau$ for each year, $2004$ to $2020$, to find evidence of a scaling phenomenon. Within the $\tau$ range where the invariance is identified ($\tau = \text{1000 min}$ to $\text{30000 min}$) -- the regime highlighted by light cyan, we observe a near-linear relationship between $\ln \tau$ and $\ln b$ suggesting a Stretched Exponential or Power law fit \footnote{Note that for very small values of $1\textrm{min} \leq \tau \leq 200\textrm{min}$, the estimated pairwise correlations aren't reliable due to Epps effect~\cite{epps}.}. We now evaluate these fits using Model Architecture Search (MAS).

\subsection*{Model Architecture Search}

\begin{figure}[h]
\centering
\includegraphics[width=1.0\textwidth]{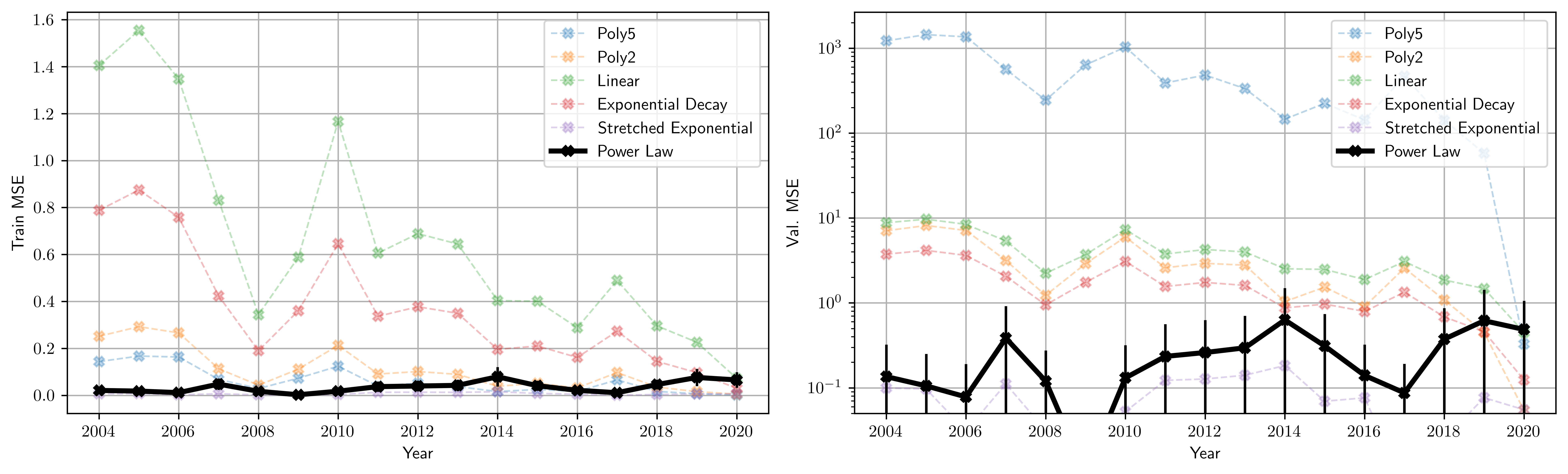}
\caption{{\bf Model Architecture Search - Training and Validation MSE:} The MSE (CI = $\sigma$ between folds) is reported as a measure-of-fit of every model for each year for $\{\tau, b(\tau)\}$ samples within the identified $\tau$ regime. Observe that the Power Law and the Stretched Exponential fits consistently reports lower validation MSE. Error bars are computed across $4$ folds of cross-validation. Polynomial models demonstrate clear signs of overfitting while the exponential model ($\beta = 1$) is only slightly worse than the best fits. A similar analysis with $T_{int} = 3$ months is presented in the Appendix Fig.~\ref{supp:mas}.}
\label{fig:mas}
\end{figure}

We consider $6$ candidate regression models to fit the $\{\tau, b(\tau)\}$ data samples from $\tau = 1000$min to $30000$min: Linear, Polynomial (degree=2,5), Exponential, Stretched Exponential, and Power Law. A $4-$fold cross-validation setup is used: For every year between $2004$ and $2020$, we fit each candidate model on $75\%$ of the samples and report the training and validation Mean Squared Error (MSE) on the remaining $25\%$. Error bars indicate the standard deviation of the MSE across the $4$ folds. In the case of log-transformed target variables, the MSE is computed in the original scale to ensure fair comparison. Fig.~\ref{fig:mas} indicates that the Power Law and Stretched Exponential models have the best fits among the candidates. When $T_{int} = 1$year, the Stretched Exponential fit is slightly better. When $T_{int} = 3$months, the Power Law fit is marginally better (see Fig.~\ref{supp:mas} in the Supporting Information).


\section*{Generative Models}
\begin{figure}[t]
\centering
\includegraphics[scale=0.30]{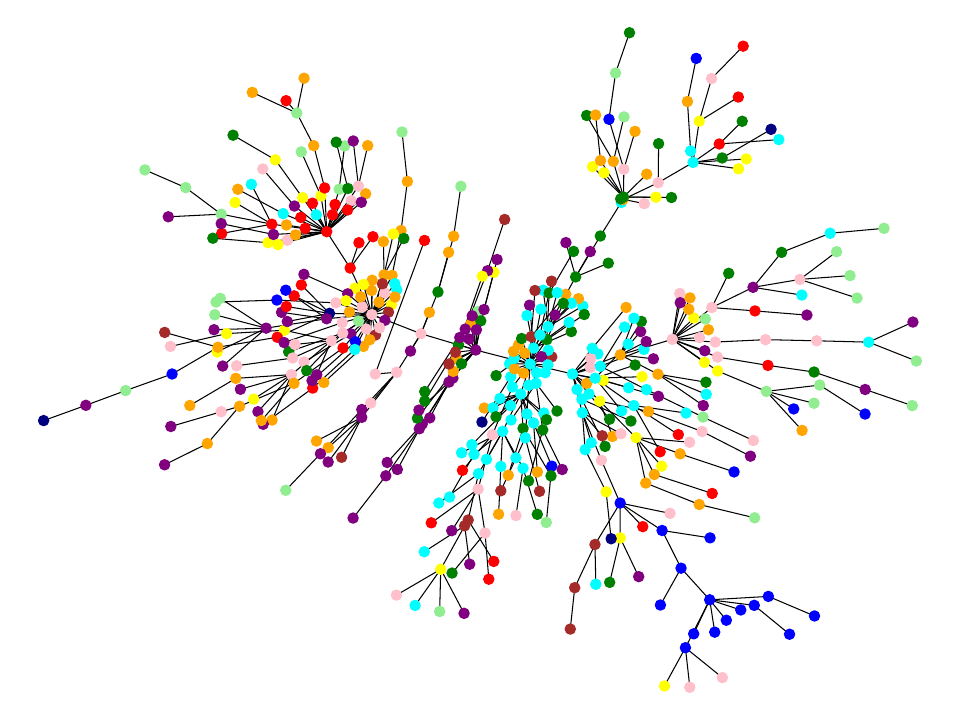}
\includegraphics[scale=0.30]{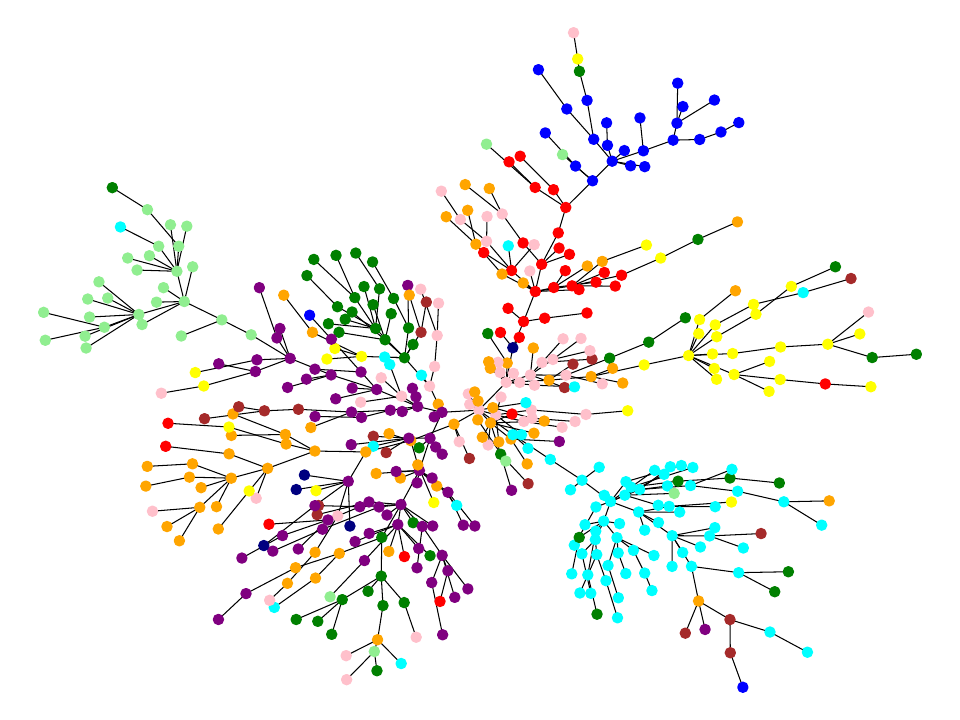}
\includegraphics[scale=0.40]{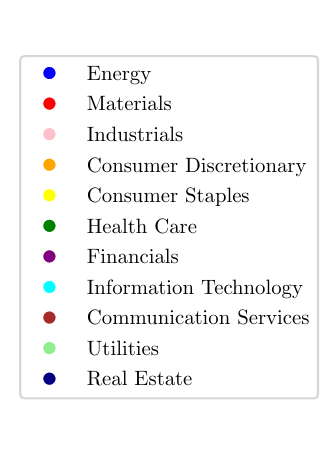}
\caption{{\bf MSTs with (left) $\tau = \text{1 min}$ and (right) $\tau = \text{1000 min}$ in 2004:} The starting and end dates are Jan 1st and Dec 31st, respectively. Each vertex is colored depending on its GICS sector. Edge weights are computed as: $d_{i, j} (\tau) \coloneqq \sqrt{2 (1 - \rho_{i, j} (\tau))}$. Observe that for larger values of $\tau$, stocks belonging to the same sector cluster together.}
\label{main_numerical_result_ensemble_00001_years_00001_mst_001_001}
\end{figure}

We have observed so far that real S\&P500 data demonstrates a functional invariance in the standardized distributions of the partial return correlations and an associated linear dependence of the precision with respect to $\tau$. Economists attempt to construct generative models to explain these results in order to better characterize the consensus-forming taking place in the stock market. A starting point -- as noted in the Introduction -- is the correlation graph of inter-connected stocks, which reveals emergent stock communities for a return horizon $\tau > \tau_0$ corresponding to industry sectors. These are shown in Fig.~\ref{main_numerical_result_ensemble_00001_years_00001_mst_001_001}.





Many generative models -- such as the single- and multi-factor models -- have been proposed to explain these interactions by quantifying the pair-wise inter-stock interactions. We show that these do not replicate the observed functional invariance and/or scaling behavior and propose a suitable replacement -- a modified Vicsek model -- that is more promising.

\subsection*{Factor models} \label{main_sec_factor_model_001}
\subsubsection*{Single Factor Model}

The conventional single-factor model~\cite{Luenberger_001} uses only the fluctuations of the market mode and individual stock prices to model the correlations of return, i.e.,
\begin{equation}
r_{i}(t)=\alpha_{i}+\beta_{i}r_{0}(t)+\xi_{i}(t)\ ,\label{eq:ofm_eq}
\end{equation}
where $r_{0}(t)$ represents the market mode describing the overall
fluctuation of the financial market. In Eq.
(\ref{eq:ofm_eq}), $\xi_{i}(t)$ is the part not included in the
market mode. In the one-factor model, $\xi_{i}(t)$ is a zero mean
Gaussian distributed time series with $\langle\xi_{i}^{2}\rangle=\epsilon_{i}^{2}$
and is independent of each other and $r_{0}(t)$. In fact, we can derive the values of the correlation coefficients
in the one-factor model: $
\rho_{ij}=\rho_{i0}\rho_{j0}$, and the residuals $\{ c_{i, j} (\tau) \}_{i, j = 1}^N = 0$ in Eq.~\eqref{main_def_nonmarket_correlation_001_001}. The standardized distribution $\tilde{p}_\tau (\cdot)$ in Eq.~\eqref{main_def_scaled_distribution_cij_001_001} reduces to the delta function ($\mu = 0$, $b(\tau) \rightarrow \infty$) violating the structure of the empirically observed correlation distribution.

\subsubsection*{Multi-Factor Model}
The MSTs of the  pairwise stock correlations clearly show clustering of stocks belonging to the same sector, and one can formulate a multi-factor model~\cite{mfm_1, mfm_2, mfm_3} wherein we supply additional parameters that correspond to the individual sectors. Since the computational models we are considering directly output returns (not the prices), one needs to introduce an additional parameter to simulate the effect of time scale $\tau$: by varying this parameter, one can control whether the market mode dominates --drowning out the effect of the sectors (as observed for small $\tau$ in real data)-- or is suppressed, allowing the sector correlations to emerge (as observed for large $\tau$).

\begin{figure}[t]
\centering
\includegraphics[scale=0.20]{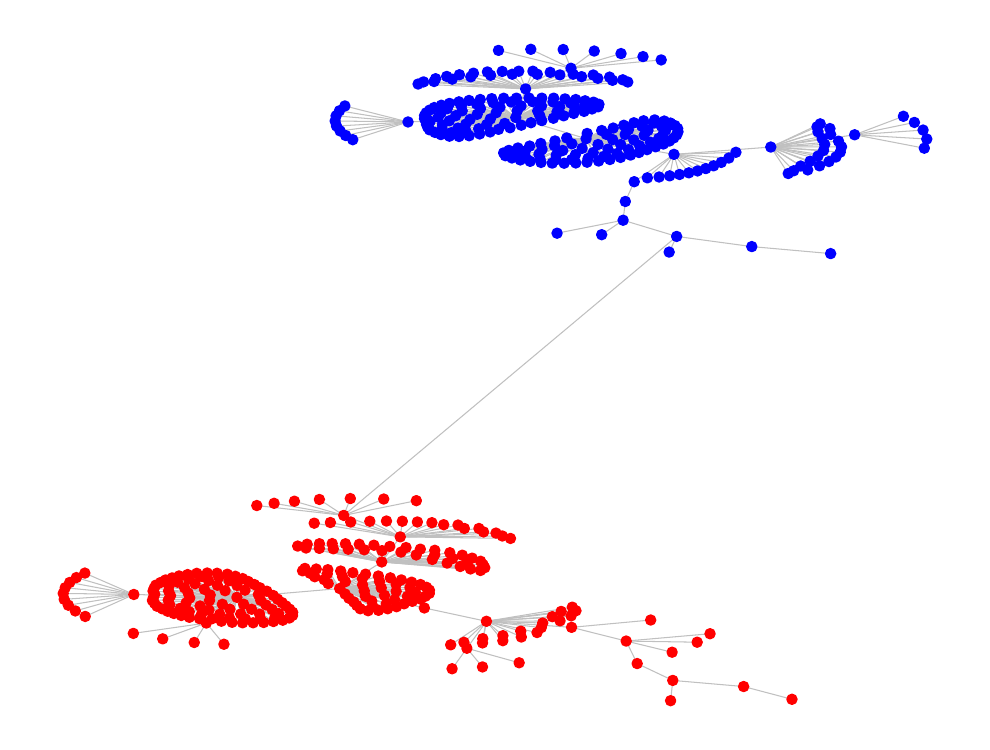}
\includegraphics[scale=0.20]{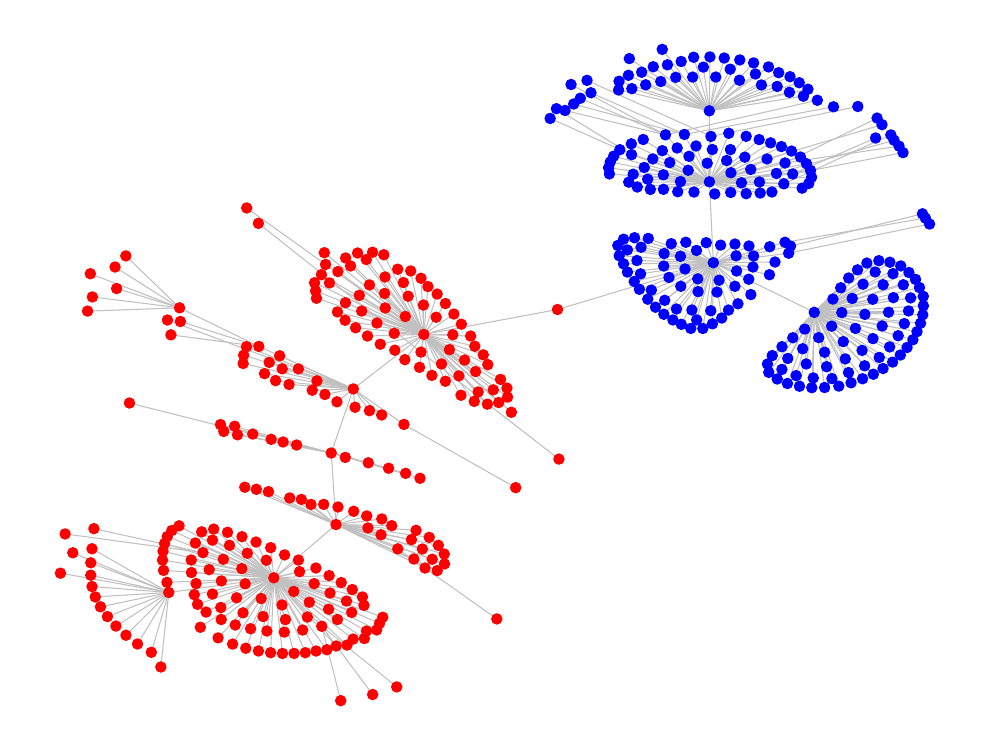}
\includegraphics[scale=0.20]{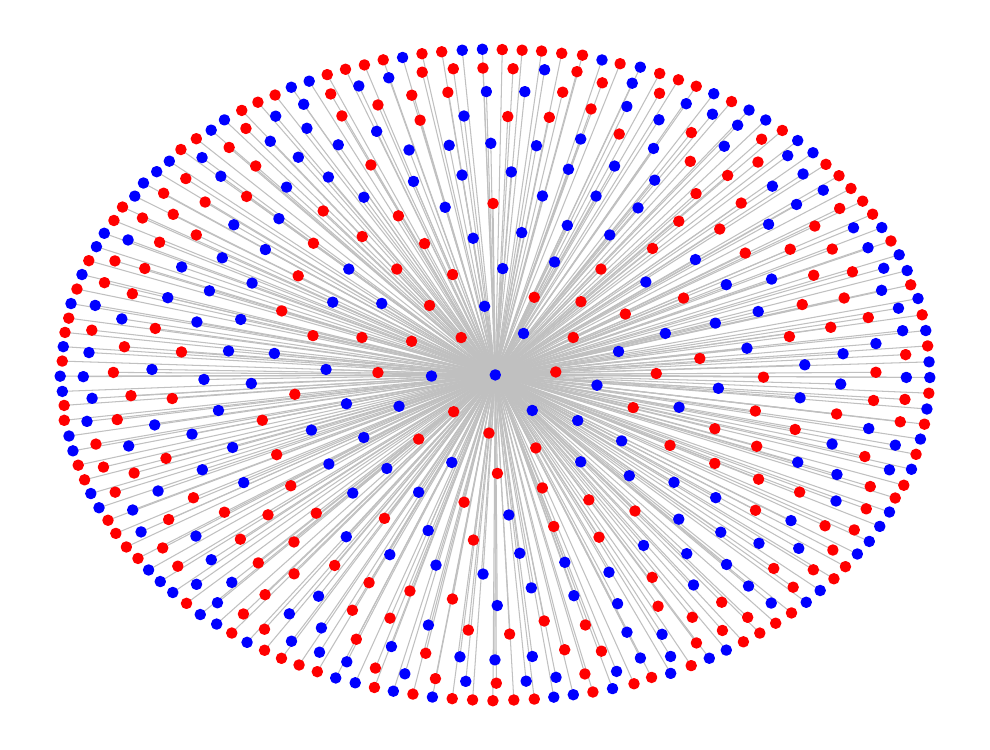}
\caption{{\bf MSTs of the multi-factor models:} The sector for each stock (node) is indicated by its color. We varied $\tilde{\eta}$: (left) $\tilde{\eta} = 0.1$, (center) $\tilde{\eta} = 10.0$, and (right) $\tilde{\eta} = 1000.0$. We set the number of steps 10000 and $\mu_{\tilde{\gamma}} = 1.0$. Observe for small $\tilde{\eta}$, the sectors are separated into distinct groups in the MST -- similar to when $\tau$ is large. As $\tilde{\eta}$ increases, the groups lose identity and merge; i.e the sector information is devalued: a similar effect to when $\tau$ is small. Precision in the edge weights is set to $1$.}
\label{main_numerical_result_MST_multi_factor_model_001_001}
\end{figure}

\begin{figure}[t]
\centering
\includegraphics[scale=0.45]{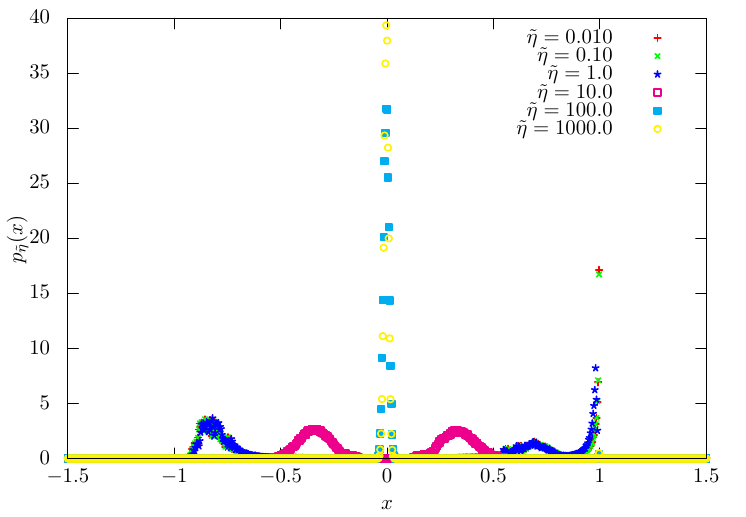}
\includegraphics[scale=0.45]{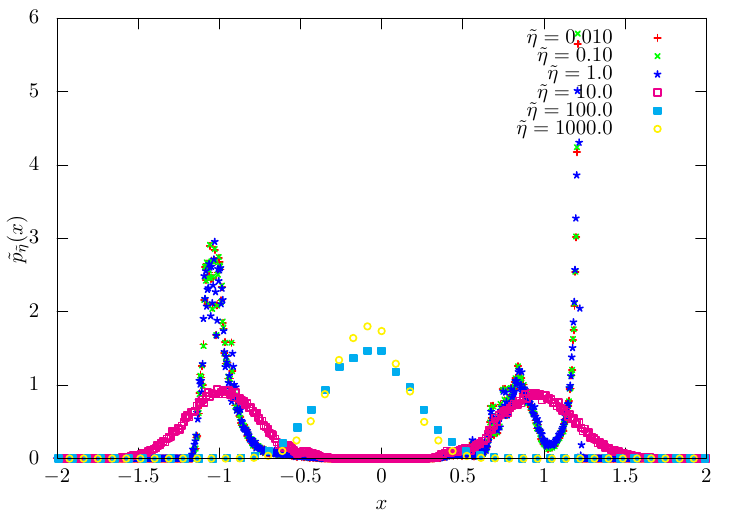}
\caption{{\bf PDFs for multi-factor models:} (a) PDF $p_{\tilde{\eta}} (x)$ in Eq.~\eqref{main_def_p_tau_c_001_001} of the multi-factor model; (b) Standardized PDF $\tilde{p}_{\tilde{\eta}} (x)$ in Eq.~\eqref{main_def_scaled_distribution_cij_001_001}. We set the number of steps 10000 and $\mu_{\tilde{\gamma}} = 1.0$.}
\label{main_numerical_result_cij_original_factor_model_001_001}
\end{figure}

Consider $K$ sectors in a market. The multi-factor model takes the form:
\begin{align}
r_i (t; \tilde{\eta}) &= \tilde{\alpha}_i + \tilde{\beta}_i r^\mathrm{mar} (t) + \sum_{k=1}^K \tilde{\gamma}_{i, k} r_k^\mathrm{sec} (t) + \tilde{\xi}_i (t; \tilde{\eta}),
\end{align}
where,
\begin{align}
  r^\mathrm{mar} (t + \Delta t) - r^\mathrm{mar} (t) &\sim \mathcal{N} (0, 0.050 \times \Delta t), \\
  r^\mathrm{sec} (t + \Delta t) - r^\mathrm{sec} (t) &\sim \mathcal{N} (0, 0.10 \times \Delta t),
\end{align}
and, for $i = 1, 2, \dots, N$,
\begin{align}
  \tilde{\xi}_i (t; \tilde{\eta}) &\coloneqq \frac{d}{dt} \tilde{\Xi}_i (t, \cdot), \\
  \tilde{\Xi}_i (t + \Delta t, t) &\coloneqq \mathcal{N} (0, \tilde{\eta}^2 \times \Delta t). \label{main_eq_Xi_001_001}
\end{align}
Here, $\mathcal{N}(\mu, \sigma^2)$ is the Gaussian perturbation function and $\tilde{\gamma}_{i, k}$ is non-zero when stock $i$ belongs to sector $k$ and $0$ otherwise. Additionally, the variance of market and sector returns are set to $0.05 \times \Delta t$, and $0.10 \times \Delta t$ respectively. Note that increasing $\tilde{\eta}$ corresponds to larger perturbations of $\tilde{\xi}_i (t)$ in successive time steps. Thus it plays the role of $1/\tau$: a large $\tilde{\eta}$ implies market-mode dominance and small $\tilde{\eta}$, sector-mode dominance (see Figs.~\ref{main_numerical_result_MST_multi_factor_model_001_001} and ~\ref{main_numerical_result_cij_original_factor_model_001_001}).


We performed the numerical simulation of the multi-factor model with $K = 2$ sectors.  We set $N = 500$ and the number of stocks in each sector as $250$.
We swept $\tilde{\eta}$ in Eq.~\eqref{main_eq_Xi_001_001} across multiple values. The other parameters are set as follows: $\mu_{\tilde{\gamma}} = 1.0$, $\tilde{\alpha}_i \sim \mathcal{N} (0.0, 1.0)$, $\tilde{\beta}_i \sim \mathcal{N} (0.0, 1.0)$, and $\tilde{\gamma}_{i, k} \sim \mathcal{N} (\mu_{\tilde{\gamma}}, 1.0)$ if stock $i$ belongs to sector $k$ and otherwise zero. In Fig.~\ref{main_numerical_result_MST_multi_factor_model_001_001}, we show the MSTs of the multi-factor models for $\tilde{\eta} = 0.1, 10.0, 1000.0$ and $\mu_{\tilde{\gamma}} = 1.0$. Observe that for small $\tilde{\eta}$, the stocks per sector belong in separate communities in the MST. As $\tilde{\eta}$ increases, the communities collapse.


In Fig.~\ref{main_numerical_result_cij_original_factor_model_001_001}, we plot the PDF $p_\tau (\cdot)$ in Eq.~\eqref{main_def_p_tau_c_001_001} of the multi-factor model and standardized PDF $\tilde{p}_\tau (\cdot)$ in Eq.~\eqref{main_def_scaled_distribution_cij_001_001}. The following dynamics are observed: (a) For small $\tilde{\eta}$ -- may correspond to large $\tau$ -- two peaks originate from two sector modes and one peak originates from the market mode; (b) for moderate $\tilde{\eta}$, the market mode dissipates and the two sector modes dominate the distribution; and (c) for large $\tilde{\eta}$, the return correlation distribution becomes random due to large perturbations. The multi-factor model explicitly uses sector affiliation as a parameter resulting in multi-modal correlation distributions. This multi-modal structure of $\tilde{p}_{\tau}(\cdot)$ results in the precision of return correlations $b(\tau)$ not scaling with $\tau$.

\subsection*{Modified Vicsek model} \label{main_sec_vicsek_like_model_001}

The Vicsek model is a generative model that can display some of the salient group  characteristics of swarming behavior, as observed in the motion patterns of flocks of birds and swarms of fish. Compared to the multi-factor model where  group assignments are provided in advance, such assignments emerge naturally in the Viscek model:  each particle in the swarm is influenced by other particles that are within a neighborhood. Based on such local-only interactions, long distance order emerges and groups of particles cluster together in their dynamical behavior, akin to sectors emerging in stock markets. 

Our model uses the standard setup ~\cite{Vicsek001} with the following modifications: (a) Consistent with the factor model setup, the predicted variable is the return $r_i(t)$; (b) Particles (individual stocks) move in $\mathbb{R}^1$ (rather than in $\mathbb{R}^2$); a stock's offset from $0$ is the return value. The proximity of one particle $i$ to another $j$ at time $t$ is the absolute value of the difference of the returns $|r_i(t) - r_j(t)|$ rather than the typical cosine distance metric used in $\mathbb{R}^2$; (c) Time steps are discretized rather than continuous. The update step is:

\begin{align}
  \alpha_i r_i (t + \Delta t) &= (\alpha_i - \beta_i \Delta t) r_i (t) \nonumber \\
  & \quad + \frac{\gamma_i}{N_{i, \delta}} \sum_{j: |r_i (t) - r_j (t)| < \delta} (r_j (t) - r_j (t - \Delta t)) \nonumber \\
  & \quad + \Xi_i (t + \Delta t, t), \label{main_eq_Vicsek_like_model_001_001}
\end{align}
where
\begin{align}
  \Xi_i (t + \Delta t, t) &\sim \mathcal{N} (0.0, \eta^2 \Delta t), \label{main_eq_Vicsek_like_model_001_002}
\end{align}
and $N_{i, \delta}$ is the number of elements $j$ that satisfy $|r_i (t) - r_j (t)| < \delta$. An extended derivation of the Vicsek update step is presented in \nameref{S2}. 

\begin{figure}[t]
\centering
\includegraphics[scale=0.25]{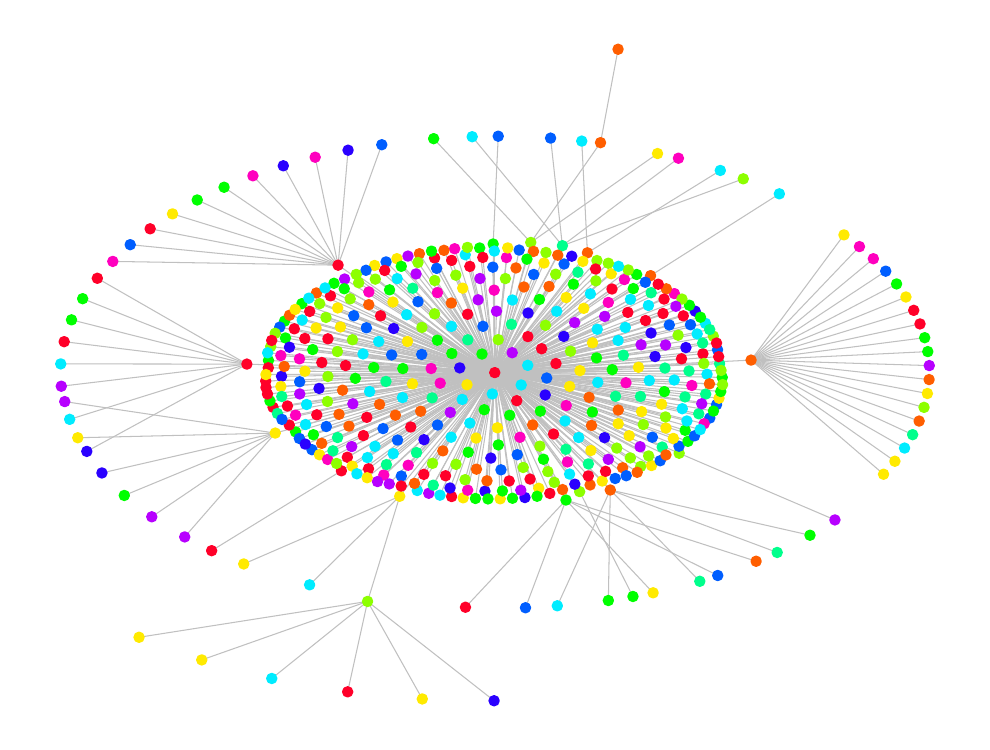}
\includegraphics[scale=0.25]{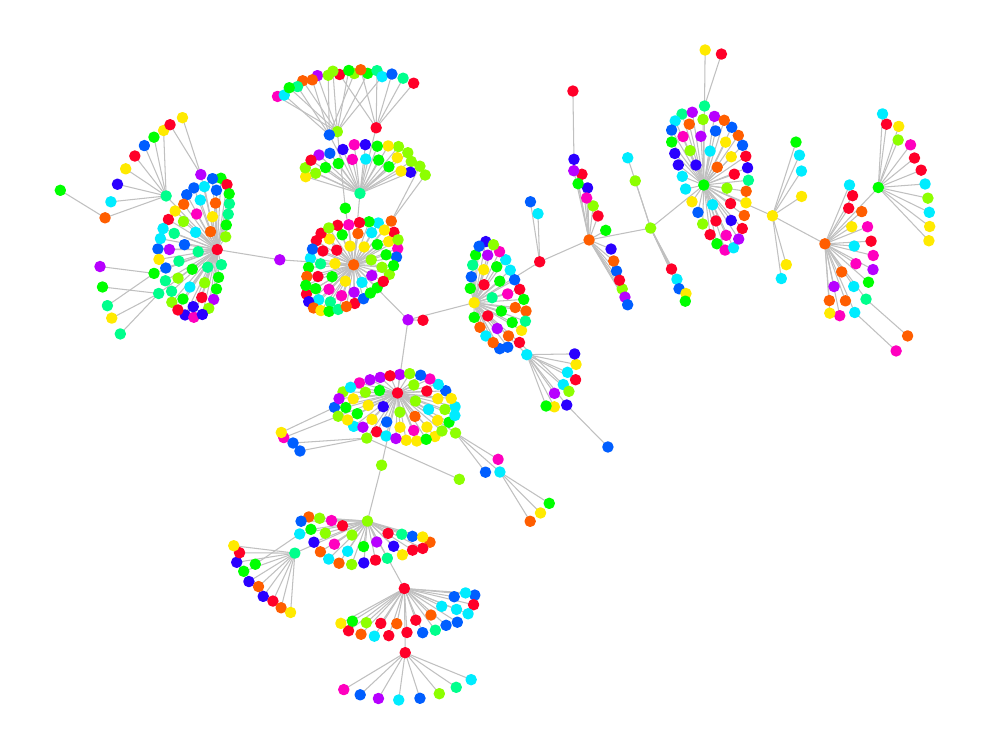}
\includegraphics[scale=0.25]{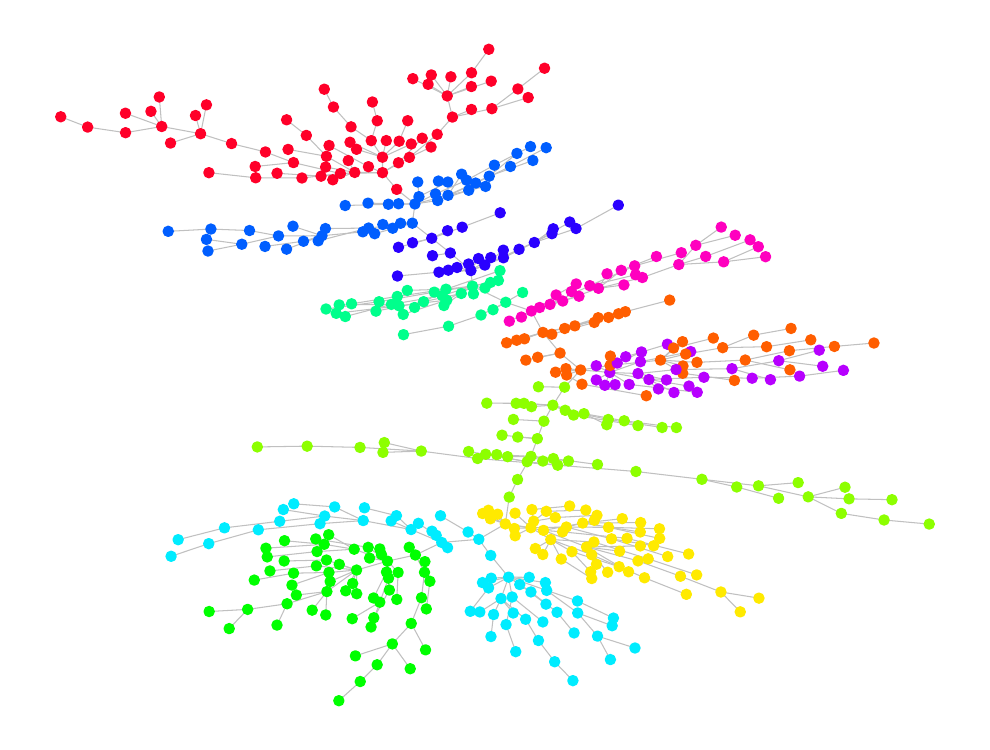}
\caption{\textbf{Minimum Spanning Trees (MSTs) obtained from the modified Vicsek generative model with varying $\eta$ values:} We generated MSTs using the modified Vicsek generative model with different values of $\eta$: (a) $\eta = 0.010$, (b) $\eta = 1.0$, and (c) $\eta = 100.0$. The number of steps was set to 10000, and $\delta$ was fixed at 0.10. For $\eta = 100.0$, the vine structure is apparent and we used these vines to define the analogs of sectors in stocks. In particular, we performed a community finding on MST~\cite{girvannewman} corresponding to the vine structure, and identified 11 communities corresponding to the number of GICS sectors. These communities indeed constitute individual vines, as shown by colored nodes in the right-most figure. Next we tracked the associated stocks as $\eta$ decreased based on the fixed $\delta$ condition. Notably, as $\eta$ decreases, the sectors collapse due to the fixed neighborhood of $\delta$, which encourages more particles (stocks) to interact with one another. This increased interaction arises as the particles experience less perturbation from $\Xi$, leading to homogeneous behavior and radial MSTs.}
\label{main_numerical_result_MST_Vicsek_like_model_001_001}
\end{figure}

\noindent \textbf{Evaluating the Vicsek model under different parameter settings of} $\delta,\eta$: We next discuss how  parameters $\delta$ (radius of influence), and $\eta$ (standard deviation of noise) -individually and collaboratively - can play  roles analogous to the return horizon parameter $\tau$ in the empirical stock price data. In particular, we analyze the dependence of the distributions of correlation of returns defined in Eq.~\eqref{main_eq_Vicsek_like_model_001_001} on $\delta$ and $\eta$.

\begin{figure}[t]
\centering

\includegraphics[width=0.48\textwidth]{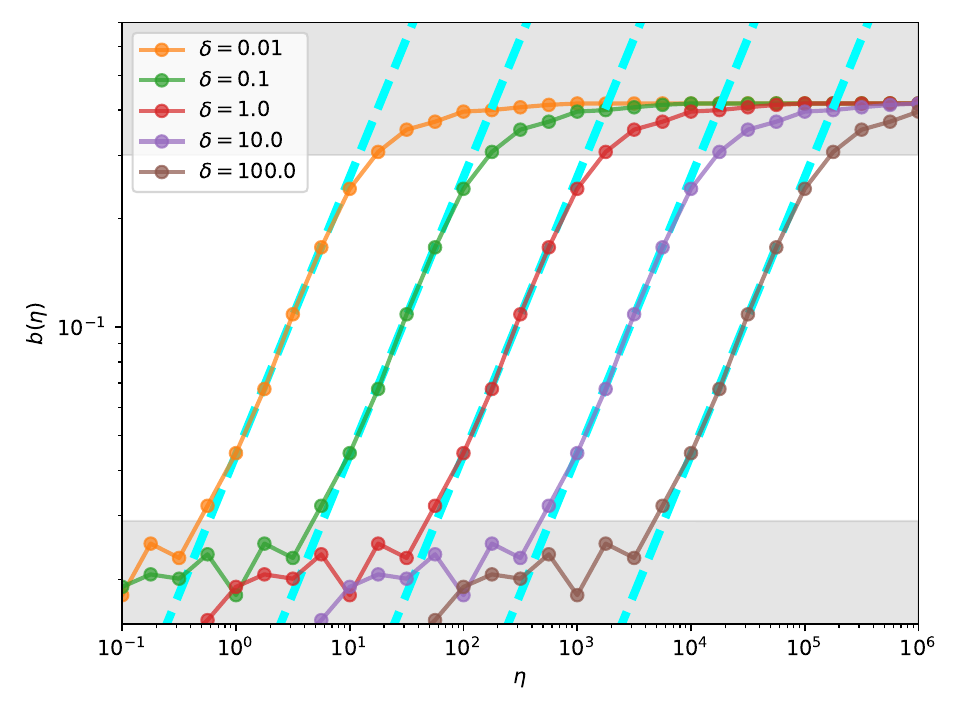}
\includegraphics[width=0.48\textwidth]{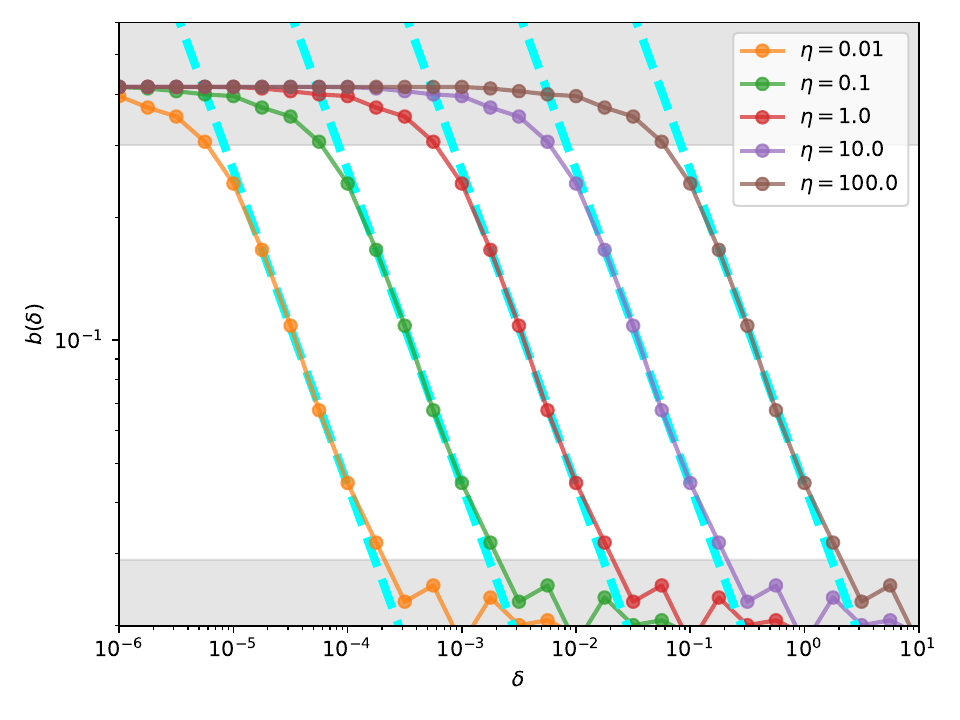}
\caption{{\bf Visualizating the dependency of $b(\cdot)$ w.r.t (a) $\eta$ and (b) $\delta$}: Observe a scaling phenomenon of $b(\cdot)$ with respect to $\eta$ and $\delta$ as the empirically observed trend in Fig.~\ref{main_numerical_result_ensemble_00001_years_00001_scaling_law_001_001}: Number of steps = $10000$. The highlighted region (in which) denotes the near-linear fit between $b(\cdot)$ and $\eta, \delta$. For different parameter settings, the near-linear fit in the log-log plot is visualized in cyan.}
\label{main_numerical_result_MST_Vicsek_like_model_001_002}
\end{figure}

\begin{itemize}[leftmargin=0.5cm]
\item \noindent \textit{Role of $\delta$}: The parameter $\delta$ plays a crucial role in determining the extent to which particles in the model are influenced by their neighbors. We anticipate that very large values of $\delta$ lead to substantial inter-particle influence, producing highly correlated return sequences, while very small $\delta$ values result in independent particle behavior with little correlation. Thus, we would expect small values of $\delta$ to lead to very small return correlations -- akin to short return horizons $\tau$ -- and as $\delta$ is increased we expect pockets of correlated returns, just as sectors emerge in the empirical data with increasing $\tau$. Indeed, as shown in Fig.\ref{main_numerical_result_MST_Vicsek_like_model_001_002}, for intermediate values of $\delta$ the precision follows a near-linear dependence in the log-log scale with respect to $\delta$ -- a scaling phenomenon.
\item \noindent \textit{Role of $\eta$}: Injected noise adds randomness to the trajectories of particles and together with the radius of influence determined by $\delta$, the noise level $\eta$ facilitates the formation of distinct pockets. In the absence of this noise and with a sufficiently wide radius of influence, particles tend to merge into a unified group, exhibiting strong correlations with each other. Visual evidence illustrating this effect can be seen in the MST structure in Fig.~\ref{main_numerical_result_MST_Vicsek_like_model_001_001}(a). As illustrated in Fig.~\ref{main_numerical_result_MST_Vicsek_like_model_001_001}(b) and ~\ref{main_numerical_result_MST_Vicsek_like_model_001_001}(c), increasing the noise factor results in the formation of communities. Of course if $\eta$ is increased further, the vine structure will disintegrate.
\end{itemize}

\textit{Indeed, $\delta$ and $\eta$ behave as duals of one another while influencing the distribution of the return correlations:}  For example, increasing the noise in particle trajectory ($\eta$) has a similar effect to decreasing each particle's radius of influence ($\delta$). Given these constraints, we look to discover a scaling effect with respect to $\delta, \eta$ and functional invariance of $\tilde{p}_\delta(\cdot), \tilde{p}_\eta(\cdot)$  for \textit{intermediate} $\delta, \eta$ values. For simulations, we set $N = 500$, $\alpha_i = \gamma_i = 1.0$ and $\beta_i = 0.05$ for $i = 1, 2, \dots, N$, and $\Delta t = 1.0$.

\begin{figure}[t]
\centering
\includegraphics[scale=0.49]{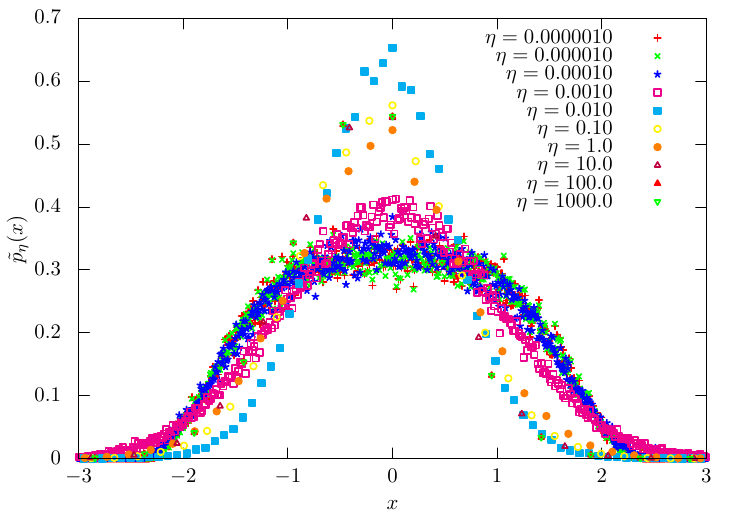}
\includegraphics[scale=0.49]{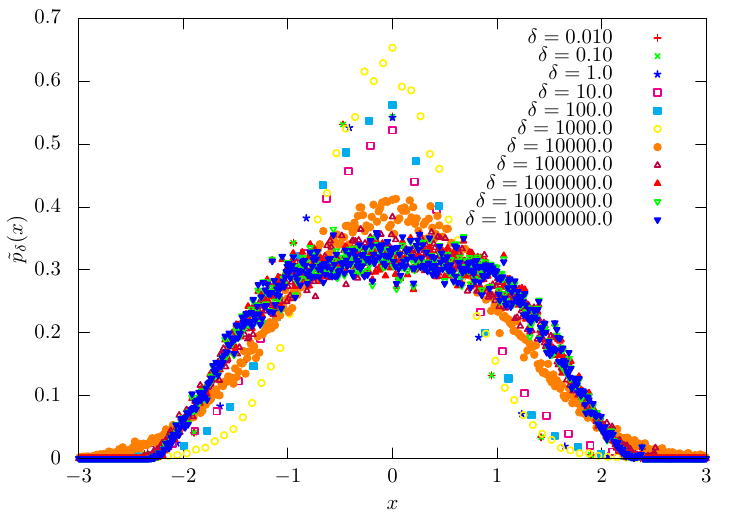}
\caption{{\bf Standardized PDF for the Vicsek model:} (a) Standardized PDF $\tilde{p}_\eta (x)$ in Eq.~\eqref{main_def_scaled_distribution_cij_001_001} of the modified Vicsek model. Note that the subscript is changed from $\tau$ to $\eta$. We set $\delta = 1.0$ and the number of steps $10000$. (b) Standardized PDF $\tilde{p}_\delta (x)$ in Eq.~\eqref{main_def_scaled_distribution_cij_001_001} of the modified Vicsek model. Note that the subscript is changed from $\tau$ to $\delta$. We set $\eta = 10.0$ and the number of steps $10000$.}
\label{main_numerical_result_cij_scaled_vicsek_like_model_001_001}
\end{figure}

\begin{itemize}[leftmargin=0.5cm]

    \item \textit{Functional form of the correlation PDFs:} In Fig. \ref{main_numerical_result_cij_scaled_vicsek_like_model_001_001}, we present the standardized correlation PDF $p_{(\cdot)} (\cdot)$ for various $\eta$ values (on the left) and $\delta$ values (on the right) (compare with the empirical result in  Fig.\ref{main_numerical_result_ensemble_00001_years_00001_cij_scaled_raw_data_001_001, main_numerical_result_ensemble_00001_years_00001_cij_scaled_raw_data_001_002}). Notably, within a finite range of $\eta$ and $\delta$, we observe that the functional form shows invariance properties similar to those observed in the empirical data.
    
    \item \textit{Scaling behavior with respect to $\eta$ and $\delta$:} In Fig.~\ref{main_numerical_result_MST_Vicsek_like_model_001_002}, we plot the relationship between the precision and each of the parameters $\eta$ (left) and $\delta$ (right) keeping the other fixed (please refer to Fig.~\ref{main_numerical_result_ensemble_00001_years_00001_scaling_law_001_001} for a comparison). 
    We observe the scaling phenomenon for intermediate values of $\eta$ and $\delta$. While at the extremes, particle trajectories are either completely uncorrelated (high $\eta$, low $\delta$) or globally correlated (low $\eta$, high $\delta$), the range in between facilitates particles to be locally correlated (akin to sectors -- see Fig.~\ref{main_numerical_result_MST_Vicsek_like_model_001_001} (right)).

\end{itemize}


\section*{Concluding Remarks}

\begin{figure}[!h]
\centering
\includegraphics[width=1.0\columnwidth]{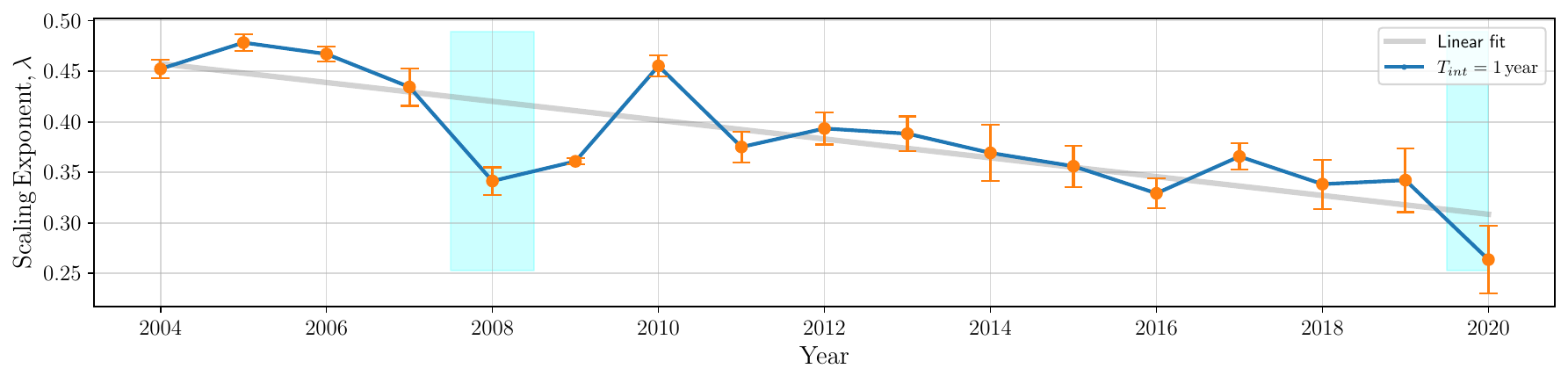}
\caption{{\bf Scaling exponent with respect to year:} Values of $\lambda$ from 2004 to 2020 with interval $T_{int} = 1$ year. The definition of $\lambda$ is given in Eq.~\eqref{main_eq_scaling_b_001_001}. Error bars are computed using 4-fold cross-validation while estimating the linear fit. In the blue highlighted regions, anomalies are observed where $\lambda$ deviates from the linear fit.}
\label{main_numerical_result_ensemble_00001_months_00001-00012_scaling_factor_001_001}
\end{figure}

In this paper, we first observe that the standardized distributions of the partial correlation of returns reaches an invariance for a finite range of $\tau$. Second, within this $\tau$ regime, we demonstrate a scaling phenomenon governing the precision of the raw distributions, $b(\tau)$, with respect to $\tau$ -- the investment horizon. We additionally review existing stochastic and generative factor models to show that they fail to model these observed emergent phenomena and propose a modified Vicsek-inspired framework that is a more promising candidate. The scaling behavior was demonstrated yearly from $2004$ to $2020$ on real stock price data sampled every minute of trading hours.


The compelling presence of such a scaling phenomenon warrants investigating the  role of the model parameters that are crucial to the fit that explains the dependence of $b(\tau)$ on $\tau$. Specifically, in the case of a Power Law fit, $\lambda$ appears as a macro-economic indicator of market health. A similar analysis on the Stretched Exponential fit -- also a good fit of the $\{\tau, b(\tau)\}$ data in Fig.~\ref{fig:mas} -- shows a similar effect with respect to the $\beta$ parameter (see Supplementary Information).

Fig.~\ref{main_numerical_result_ensemble_00001_months_00001-00012_scaling_factor_001_001} plots the scaling exponent across the $17$ years of evaluation. The figure shows that $\lambda$'s exhibit inter-annual variations, portraying a distinct linear decline from past to present, characterized by intriguing anomalies (highlighted in cyan). We seek to make sense of this trend and interpret its significance.

\noindent \textit{Setup:} Recall that the standard deviation $\sigma$ of the return correlations is proportional to $\tau^{\lambda}$. To quantify the change in standard deviation as we transition from a short-term ($\tau_I$) to a long-term ($\tau_F$) investment horizon, we introduce a novel metric defined as follows:
$$
R_\sigma (\tau_I, \tau_F) = \frac{\sigma(\tau_F) - \sigma(\tau_I)}{\sigma(\tau_F)},
$$
where $\sigma(\tau_F)$ and $\sigma(\tau_I)$ represent the standard deviations corresponding to the long-term ($\tau_F$) and short-term ($\tau_I$) investment horizons, respectively. This measure captures the fractional increase in the standard deviation from the short-term to the long-term. A large $R_\sigma$ indicates that the standard deviation of the return correlations in the short-term are much smaller than in the long-term. A small $R_\sigma$ suggests that the short- and long-term investment horizons look statistically similar.

Using the scaling law, we get:
$$
R_\lambda (\tau_I, \tau_F) = \frac{\tau_F^\lambda - \tau_I^\lambda}{\tau_F^\lambda} = 1 - \left( \frac{\tau_I}{\tau_F}\right)^\lambda,
$$
where $\frac{\tau_I}{\tau_F} \in (0, 1)$. Referring back to Fig.~\ref{main_numerical_result_ensemble_00001_months_00001-00012_scaling_factor_001_001}, we observe empirically that $\lambda \in (0,1)$. 

Therefore, $R_\lambda$ is an increasing function in $\lambda$. Given that we have noticed a consistent \textit{decrease} in the value of $\lambda$ over the years, our focus now shifts to understanding the implications of a corresponding declining trend in $R_\lambda$ across years:

\begin{itemize}[leftmargin=0.5cm]
\item \textit{Market Maturity:} We first consider the y-intercepts depicted in Fig.~\ref{main_numerical_result_ensemble_00001_years_00001_scaling_law_001_001}. Specifically, the values of $\sigma(\tau_I)$ (which is $1/b(\tau_I)$) demonstrate a consistent and gradual increase from past to present, while $\sigma(\tau_F)$ remains relatively constant across the same period. Consequently, the decreasing trend in $R_\lambda$ suggests that $\sigma(\tau_I)$ gets closer to $\sigma(\tau_F)$ every successive year. In more general terms, \textit{the communities of stocks observed over longer investment horizon in earlier years  appear in shorter time horizons in later ones, a sign that investors are becoming increasingly efficient and adept at identifying stock return patterns} - a sign of market maturity.

\item \textit{Global Financial Crises:} We now consider the cyan-colored windows in Fig.~\ref{main_numerical_result_ensemble_00001_months_00001-00012_scaling_factor_001_001} corresponding to two recent global crises -- subprime mortgage crisis in 2008 and the COVID-19 pandemic in 2020. In these cases, $\lambda$ dips significantly below the linear fit. As markets stabilized post the 2008 crisis, the $\lambda$ values rebound to the linear trend. Since our data stops at 2020, it remains to be seen whether a similar rebound will take effect.
\end{itemize}

In summary, the discovery of such scaling phenomena and its associated summary statistics in the partial correlations of stock price returns adds to a growing body of work in macro-economic modeling. By extending the qualitative observations of the variations in MST structure to the correlations at large in a quantifiable manner, we demonstrate one robust path to probe market health based on collective dynamics.

\section*{Supporting Information}

\paragraph*{S1 Appendix.}
\label{S10}
{\bf Complementary Cumulative Distribution Functions (CCDF) of the Returns as a function of $\tau$ and year} 

\begin{figure}
    \centering
    \includegraphics[width=\textwidth]{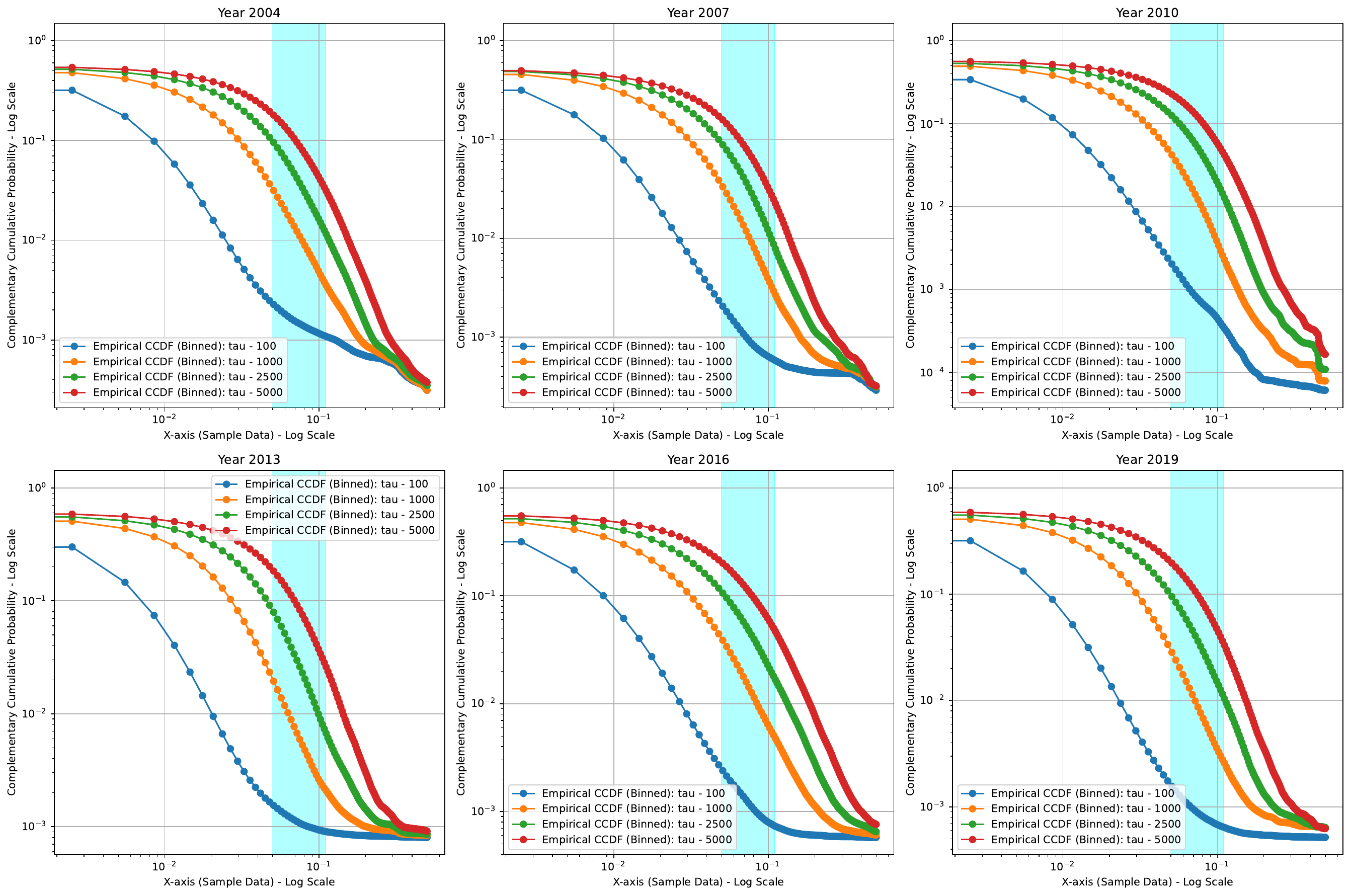}
    \caption{\textbf{Empirically estimated CCDFs of returns (+ve values only)}: The CCDF of returns computed for different return horizons $\tau$ and years are presented. In accordance to prior work, power laws are observed for a finite range of $\tau$ (marked in cyan).}
    \label{fig:ccdf}
\end{figure}

In Fig.~\ref{fig:ccdf}, we demonstrate that the CCDF of the returns follows a power law (scaling regime highlighted in cyan) similar to those observed in prior work (see Sec. Our Contributions).

\paragraph*{S2 Appendix.}
\label{S9}
{\bf The expected value (mean) of the partial correlations $\mathbb{E}(c_{ij})$ does not scale as a function of $\tau$}

\begin{figure}[h]
    \centering
    \includegraphics[width=0.6\textwidth]{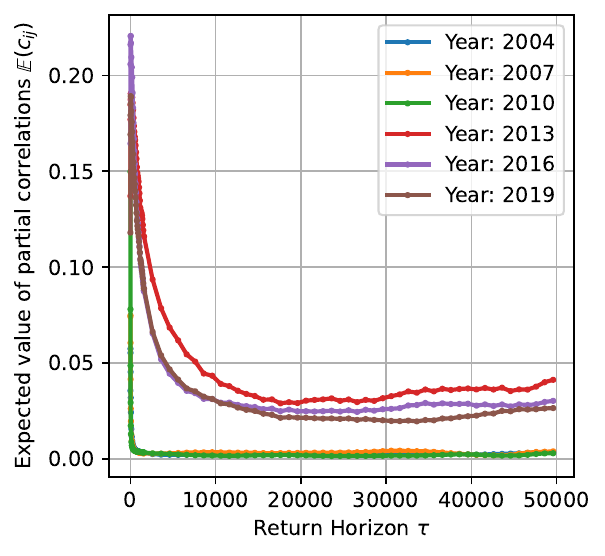}
    \caption{\textbf{Expected value of the partial correlations as a function of $\tau$ and across years}: The mean value of $c_{ij}$ as a function of $\tau$ is presented. While the standard deviation shows scaling properties, we do not observe a similar relationship governing the means.}
    \label{fig:means}
\end{figure}

In Fig.~\ref{fig:means}, we demonstrate that the mean value of the partial correlations -- as opposed to the standard deviation -- of the returns does not scale as a function of $\tau$.

\paragraph*{S3 Appendix.}
\label{S3}
{\bf Interpretability and emergent properties of the Stretched Exponential fit} \label{supp:sefit}

\begin{figure}[h]
\centering
\includegraphics[width=1.0\textwidth]{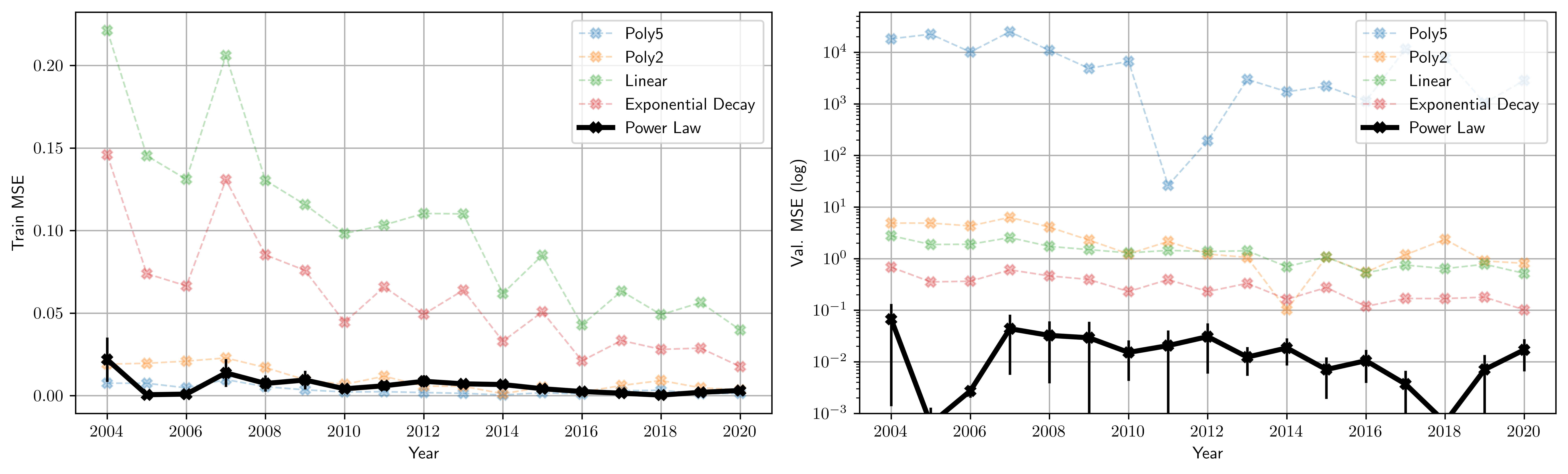}
\caption{{\bf $T_{int} = 3$ months - Model Architecture Search: Training and Validation MSE} The MSE (CI = $\sigma$) is reported as a measure-of-fit of every model for each year. Observe that the Power Law and the Stretched Exponential fits consistently reports lower validation MSE barring $2014$. Error bars are computed across $4$ folds of cross-validation. Polynomial models demonstrate clear signs of overfitting while the exponential model is only slightly worse in fit compared to the power law. These results are similar to the case with $T_{int} = 1$ year (see Fig.~\ref{fig:mas}).}
\label{supp:mas}
\end{figure}

In Figs.~\ref{fig:mas},\ref{supp:mas}, we observed that in addition to the Power Law fit, the Stretched Exponential also explains the dependence of $b(\tau)$ with respect to $\tau$. There is one extra model parameter ($3$ vs. $2$ -- see Eqs.\eqref{main_eq_scaling_b_001_001},\eqref{main_eq_scaling_b_001_002}). Similar to the effect of $\lambda$ in the case of the Power Law, in this section, we evaluate whether the stretching parameter $\beta$ in the Stretched Exponential fit can act as a market indicator.

\begin{figure}[h]
\centering
\includegraphics[width=1.0\textwidth]{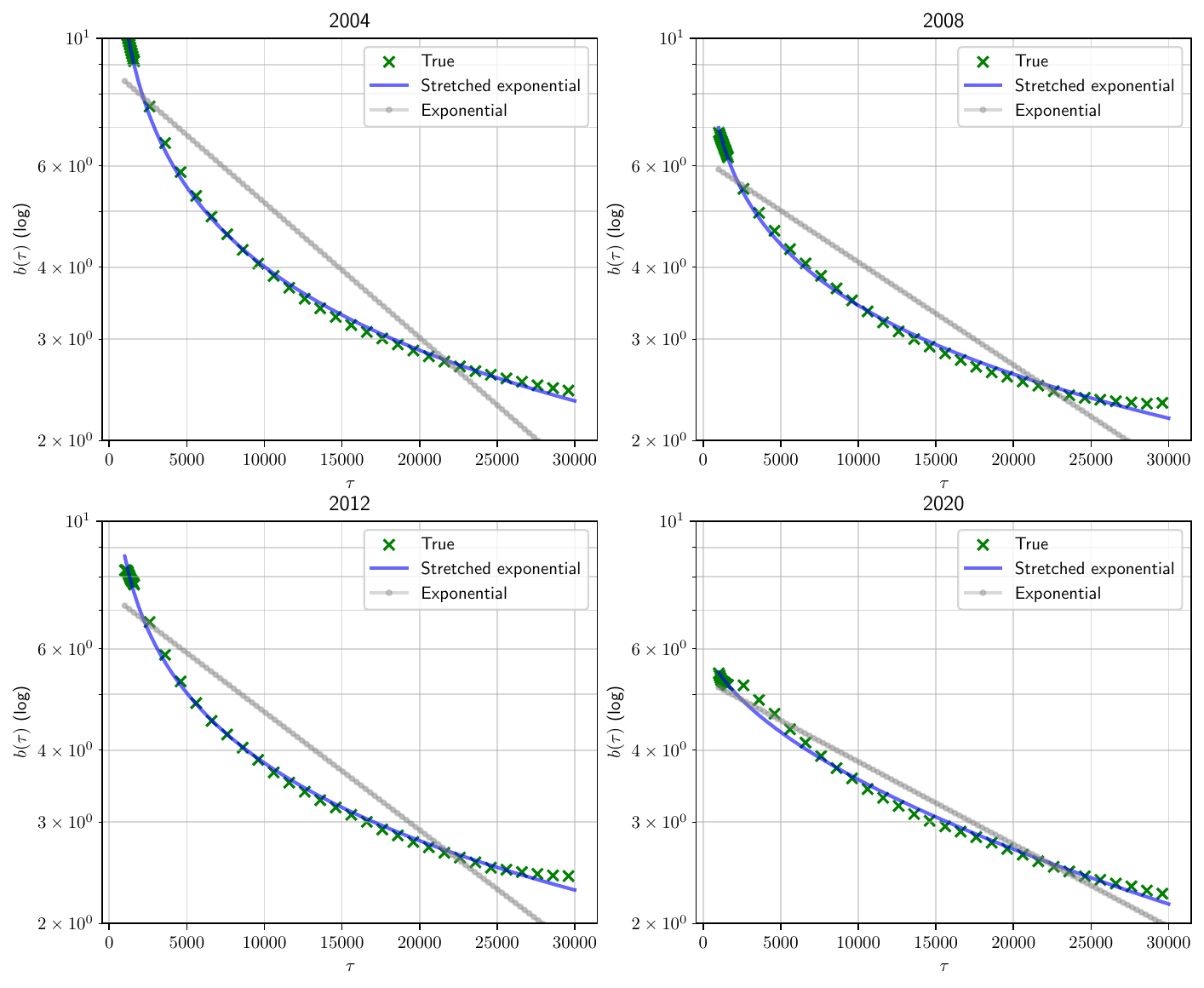}
\caption{\textbf{Plotting the dependence of $b(\tau)$ with respect to $\tau$ as a stretched exponential:} Four log-linear plots (-X-) representing the relationship between $b(\tau)$ and $\tau$ are presented for years $2004, 2008, 2012, 2020$. The stretched exponential (-, green) and exponential (grey) fits are shown. Notice that for later years, the stretched exponential fit and the exponential fit are closer to each other.}
\label{supp:fig:sefit}
\end{figure}

To get a better sense of the behavior of the stretched exponential, in Fig.~\ref{supp:fig:sefit}, we plot the $b(\tau) - \tau$ dependence in linear-log scale across $4$ years: $2004, 2008, 2012, 2020$. Both an exponential fit -- $\beta = 1$, linear in the linear-log scale -- and the best stretched-exponential ($\beta < 1$) fits are plotted. As corroborated by the MAS experiments (see Figs.~\ref{fig:mas},\ref{supp:mas}), the stretched exponential is a significantly better fit.

\begin{figure}[h]
\centering
\includegraphics[width=1.0\textwidth]{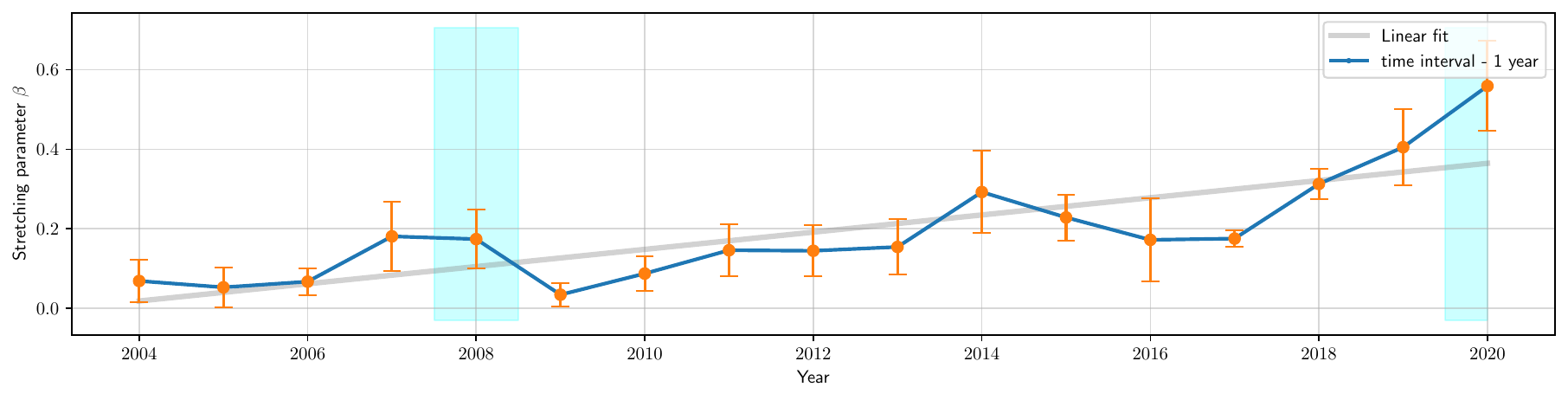}
\caption{\textbf{Stretching parameter $\beta$ with respect to year:} Values of $\beta$ from $2004$ to $2020$ with interval $T_{int} = 1$ year are plotted above along with a linear trendline fit. Error bars are generated using 4-fold cross-validation (see Fig.~\ref{main_numerical_result_ensemble_00001_months_00001-00012_scaling_factor_001_001}). Years marked with anomalies are identified in the cyan-shaded regions.}
\label{supp:beta}
\end{figure}

$\beta$ appears to be a parameter that changes year-to-year. For year $2020$, the exponential and stretched exponential fits are more similar to one another ($\beta \rightarrow 1$) than in $2004$. This motivates investigating whether there is a general trend in the $\beta$ values, similar to the $\lambda$'s presented in Fig.~\ref{main_numerical_result_ensemble_00001_months_00001-00012_scaling_factor_001_001}. In Fig.~\ref{supp:beta}, we show a similar plot.

First we explain the increasing trend: As $\beta$ increases (tends to $1$), the corresponding Stretched Exponential model fit flattens such that the short-term (small $\tau$) and long-term (large $\tau$) investment horizons are similarly effective. This can be confirmed by the y-intercepts in Fig.~\ref{supp:fig:sefit}. 

Note that this interpretation is similar to when $\lambda$ is small in the Power Law fit. In fact, there are also corresponding anomalies (marked in cyan) for the years when the market was volatile -- $2008, 2020$. Observe that for these years, the $\beta$ values significantly overshoot the linear trend (beyond an error margin; 4-fold cross validation).

\paragraph*{S4 Appendix.}
\label{S1_Exp_3}
{\bf Experiments with reduced $T_{int} = 3$ months (Q2: Apr. 1 to June 30)} \label{supp:experiments}

\begin{figure}[!h]
\centering
\includegraphics[width=0.70\columnwidth]{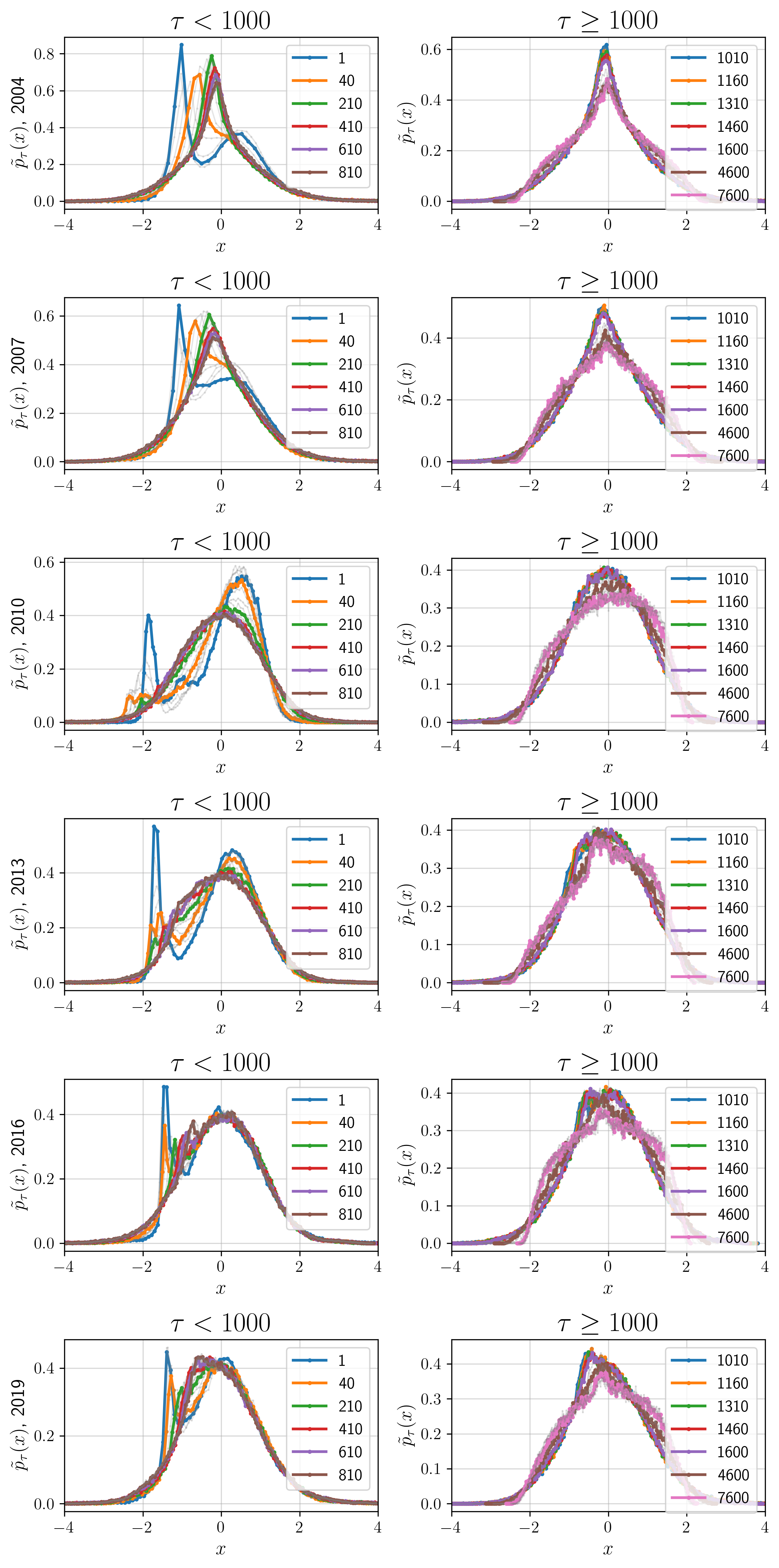}
\caption{{\bf $T_{int} = 3$ months - Qualitatively demonstrating the stability of the functional form for some $\tau > \tau_0$}: Standardized PDF $\tilde{p}_\tau (\cdot)$ in Eq.~\eqref{main_def_scaled_distribution_cij_001_001} visualized for $5$ years - 1 year per row: (a) Left: $\tau = \text{1 min}$ to $\text{1000 min}$ and (b) Right: $\tau = \text{1000 min}$ to $\text{10000 min}$. As $\tau$ exceeds $\text{1000 min}$, the shape of $\tilde{p}_\tau (\cdot)$ takes a more stable form (see Fig.~\ref{main_numerical_result_ensemble_00001_years_00001_cij_scaled_raw_data_001_001, main_numerical_result_ensemble_00001_years_00001_cij_scaled_raw_data_001_002} for the analysis with $T_{int} = 1$ year).}
\label{supp:functional_form}
\end{figure}

\begin{figure}[h]
\centering

\includegraphics[width=1.0\columnwidth]{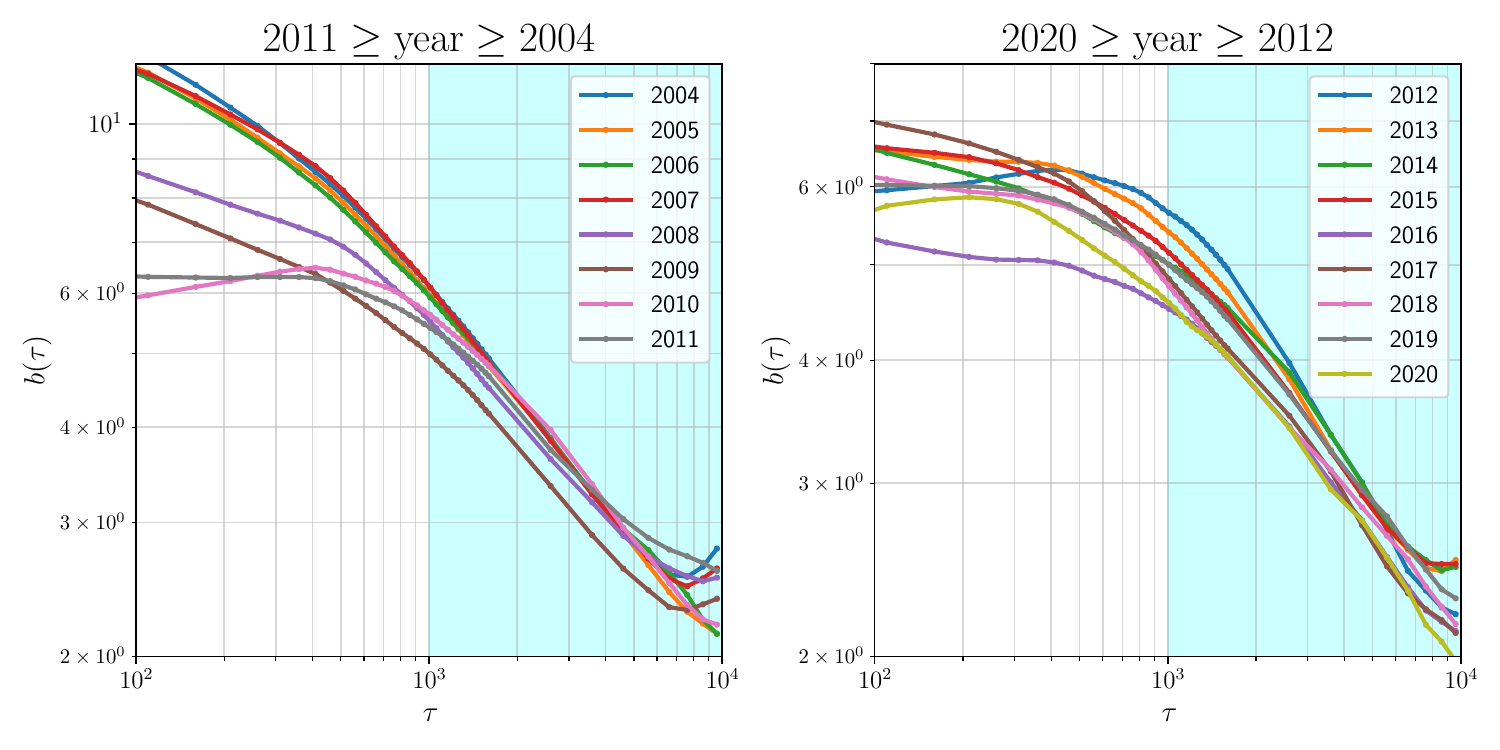}
\caption{{\bf $T_{int} = 3$ months - The power law (a) from 2004 to 2011; and (b) from 2012 to 2020} From $\tau = \text{1000 min}$ to $\text{10000 min}$ (the region highlighted by light cyan), observe the near-linear relationship between $\ln \tau$ and $\ln b$ similar to the case with $T_{int} = 1$ year (see Fig.~\ref{main_numerical_result_ensemble_00001_years_00001_scaling_law_001_001} in the main text).}
\label{supp:scaling_law}
\end{figure}

In this section, we demonstrate that (a) the functional form of the standardized PDFs $\hat{p}_{\tau}(x)$ is stable for some $\tau > \tau_0$ when $T_{int} = 3$ months between Apr. 1 and June 30 (Q2), and (b) the scaling phenomenon is once again observed between $b(\tau)$ and $\tau$. Recall that $T_{int}$ is the integration window that is used to compute the correlation in Eq.~\eqref{main_def_correlation_ij_001_001}. We accordingly decrease the maximum investment horizon from $50000$ minutes to $10000$ minutes since $\tau < T_{int}$. Below we provide $3$ figures corresponding to (a) the functional invariance (Fig.~\ref{supp:functional_form}), (b) the scaling phenomenon (Fig.~\ref{supp:scaling_law}), and (c) model architecture search (Fig.~\ref{supp:mas}). Note their similarity to those presented in the main text.

\paragraph*{S5 Appendix.}
\label{S2}
{\bf Deriving the Modified Vicsek model} \label{supp:vicsek_main}

We first review the Vicsek model~\cite{Vicsek001}. Let us consider a two-dimensional system of $N$ kinetic particles. We denote, by $\vec{x}_i (t)$ and $\vec{v}_i (t)$, the position and velocity of particle $i$ at time $t$.
The position of $i$ is updated by
\begin{align}
  \vec{x}_i (t + \Delta t) &= \vec{x}_i (t) + \vec{v}_i (t) \Delta t. \label{supp_eq_Vicsek_model_001_001}
\end{align}
The amplitude of the velocity of each particle is constant: $\| \vec{v}_i (t) \|_\mathrm{F} = v_\mathrm{abs}$.
But the direction of $\vec{v}_i (t)$ is update by
\begin{align}
  \vec{v}_i (t) &= v_\mathrm{abs}
  \begin{bmatrix}
    \cos \theta_i (t) \\
    \sin \theta_i (t)
  \end{bmatrix}, \label{supp_eq_Vicsek_model_001_002}
\end{align}
and
\begin{align}
  \theta_i (t + \Delta t) &= \frac{1}{N_{i, \delta}} \sum_{j: \| \vec{x}_i (t) - \vec{x}_j (t) \|_\mathrm{F} < \delta} \theta_j (t) + \Xi_i (t), \nonumber \\
  \Xi_i (t) &\sim [- \eta / 2, \eta / 2], \label{supp_eq_Vicsek_model_001_003}
\end{align}
where $N_{i, \delta}$ is the number of elements $j$ that satisfy $\| \vec{x}_i (t) - \vec{x}_j (t) \|_\mathrm{F} < \delta$.
Instead of the uniform distribution, we can use the normal distribution for $\Xi_i (t)$:
\begin{align}
  \Xi_i (t) &\sim \mathcal{N} \bigg( 0, \frac{\eta^2}{12} \bigg).
\end{align}

\noindent \textbf{Continuous version of the Vicsek model:} The continuous variant of Eqs.~\eqref{supp_eq_Vicsek_model_001_001}, \eqref{supp_eq_Vicsek_model_001_002}, and \eqref{supp_eq_Vicsek_model_001_003}, takes the form
\begin{align}
  \frac{d}{dt} \vec{x}_i (t) &= \vec{v}_i (t), \label{supp_eq_continuous_Vicsek_model_001_001}
\end{align}
where
\begin{align}
  \vec{v}_i (t) &= v_\mathrm{abs}
  \begin{bmatrix}
    \cos \theta_i (t) \\
    \sin \theta_i (t)
  \end{bmatrix}, \label{supp_eq_continuous_Vicsek_model_001_002} \\
  \dot{\theta}_i (t) &= - \alpha_i \theta_i (t) + \frac{\beta_i}{N_{i, \delta}} \sum_{j: \| \vec{x}_i (t) - \vec{x}_j (t) \|_\mathrm{F} < \delta} \theta_j (t) + \xi_i (t). \label{supp_eq_continuous_Vicsek_model_001_003}
\end{align}
Taking the time-derivative of Eq.~\eqref{supp_eq_continuous_Vicsek_model_001_001}, we can transform it into
\begin{align}
  \dot{\vec{v}}_i (t) &= v_\mathrm{abs}
  \begin{bmatrix}
    - \sin \theta_i (t) \\
    \cos \theta_i (t)
  \end{bmatrix}
  \dot{\theta}_i (t) \\
  &= v_\mathrm{abs}
  \begin{bmatrix}
    - \sin \theta_i (t) \\
    \cos \theta_i (t)
  \end{bmatrix}
  \bigg( - \alpha_i \theta_i (t) + \frac{\beta_i}{N_{i, \delta}} \sum_{j: \| \vec{x}_i (t) - \vec{x}_j (t) \|_\mathrm{F} < \delta} \theta_j (t) + \xi_i (t) \bigg). \label{supp_eq_continuous_Vicsek_model_003_001}
\end{align}

Again, we discretize the above equations, Eqs.~\eqref{supp_eq_continuous_Vicsek_model_001_001}, \eqref{supp_eq_continuous_Vicsek_model_001_002}, and \eqref{supp_eq_continuous_Vicsek_model_001_002}:
\begin{align}
  \vec{x}_i (t + \Delta t) &= \vec{x}_i (t) + \vec{v}_i (t) \Delta t, \label{supp_eq_continuous_Vicsek_model_002_001}
\end{align}
where
\begin{align}
  \vec{v}_i (t) &= v_\mathrm{abs}
  \begin{bmatrix}
    \cos \theta_i (t) \\
    \sin \theta_i (t)
  \end{bmatrix}, \label{supp_eq_continuous_Vicsek_model_002_002}
\end{align}
and
\begin{align}
  \theta_i (t + \Delta t) &= \theta_i (t) + \bigg( - \alpha_i \theta_i (t) + \frac{\beta_i}{N_{i, \delta}} \sum_{j: \| \vec{x}_i (t) - \vec{x}_j (t) \|_\mathrm{F} < \delta} \theta_j (t) + \xi_i (t) \bigg) \Delta t \nonumber \\
  &= (1 - \alpha_i \Delta t) \theta_i (t) + \frac{\beta_i \Delta t}{N_{i, \delta}} \sum_{j: \| \vec{x}_i (t) - \vec{x}_j (t) \|_\mathrm{F} < \delta} \theta_j (t) + \xi_i (t) \Delta t. \label{supp_eq_continuous_Vicsek_model_002_003}
\end{align}
To recover the original equation of $\theta_i (t)$, Eq.~\eqref{supp_eq_Vicsek_model_001_003}, we need to set $\alpha_i = \frac{1}{\Delta t}$ and $\beta_i = \frac{1}{\Delta t}$.
To summarize the above equations, Eqs.~\eqref{supp_eq_continuous_Vicsek_model_002_001}, \eqref{supp_eq_continuous_Vicsek_model_002_002}, and \eqref{supp_eq_continuous_Vicsek_model_002_003}, we get
\begin{align}
  \vec{v}_i (t + \Delta t) &= \vec{v}_i (t) + v_\mathrm{abs}
  \begin{bmatrix}
    - \sin \theta_i (t) \nonumber \\
    \cos \theta_i (t)
  \end{bmatrix}
  \bigg( - \alpha_i \theta_i (t) + \frac{\beta_i}{N_{i, \delta}} \sum_{j: \| \vec{x}_i (t) - \vec{x}_j (t) \|_\mathrm{F} < \delta} \theta_j (t) + \xi_i (t) \bigg) \Delta t \\
  &= v_\mathrm{abs}
  \begin{bmatrix}
    \cos \theta_i (t) \nonumber \\
    \sin \theta_i (t)
  \end{bmatrix}
  + v_\mathrm{abs}
  \begin{bmatrix}
    - \sin \theta_i (t) \nonumber \\
    \cos \theta_i (t)
  \end{bmatrix}
  \bigg( - \alpha_i \theta_i (t) + \\
  & \hspace{35pt} \frac{\beta_i}{N_{i, \delta}} \sum_{j: \| \vec{x}_i (t) - \vec{x}_j (t) \|_\mathrm{F} < \delta} \theta_j (t) + \xi_i (t) \bigg) \Delta t.
\end{align}

\noindent \textbf{Deriving the modified vicsek model:} By introducing the interaction terms in Eq.~\eqref{supp_eq_continuous_Vicsek_model_003_001} into free particles described by the Langevin equation, we propose the following model: \label{supp_sec_Vicsek_like_model_001_001}
\begin{subequations}
\begin{align}
  \dot{r}_i (t) &= v_i (t), \\
  m_i \dot{v}_i (t) &= - \alpha_i v_i (t) - \beta_i r_i (t) + \frac{\gamma_i}{N_{i, \delta}} \sum_{j: |r_i (t) - r_j (t)| < \delta} v_j (t) + \xi_i (t),
\end{align} \label{supp_eq_Vicsek_like_model_001_001}%
\end{subequations}
where $N_{i, \delta}$ is the number of elements $j$ that satisfy $|r_i (t) - r_j (t)| < \delta$ and
\begin{align}
  \xi_i (t) &\coloneqq \frac{d}{dt} \Xi_i (t, \cdot).
\end{align}
Here we call Eq.~\eqref{supp_eq_Vicsek_like_model_001_001} the modified Vicsek model.
Note that $v_i (t)$ is included in $\sum_{j: |r_i (t) - r_j (t)| < \delta} v_j (t)$ if $\delta > 0$ and the summation is zero if $\delta = 0$.
Here, for $t \ge t'$, $\Xi (t, t')$ obeys
\begin{align}
  \Xi_i (t, t') &\sim \mathcal{N} (0, \eta_i^2 (t - t')).
\end{align}
Note that, in the case of the Langevin equation, we have $\eta_i^2 = 2 \gamma k_\mathrm{B} T$.

For numerical simulations, we simplify Eq.~\eqref{supp_eq_Vicsek_like_model_001_001}.
Unifying the two equations, Eq.~\eqref{supp_eq_Vicsek_like_model_001_001}, we obtain
\begin{align}
  m_i \ddot{r}_i (t) &= - \alpha_i \dot{r}_i (t) - \beta_i r_i (t) + \frac{\gamma_i}{N_{i, \delta}} \sum_{j: |r_i (t) - r_j (t)| < \delta} \dot{r}_j (t) + \xi_i (t).
\end{align}
Taking the overdamped limit $m_i \to 0$, we get
\begin{align}
  \alpha_i \dot{r}_i (t) &= - \beta_i r_i (t) + \frac{\gamma_i}{N_{i, \delta}} \sum_{j: |r_i (t) - r_j (t)| < \delta} \dot{r}_j (t) + \xi_i (t). \label{supp_eq_Vicsek_like_model_003_001}
\end{align}

By discretizing Eq.~\eqref{supp_eq_Vicsek_like_model_003_001}, we have
\begin{align}
  \alpha_i r_i (t + \Delta t) &= (\alpha_i - \beta_i \Delta t) r_i (t) + \frac{\gamma_i}{N_{i, \delta}} \sum_{j: |r_i (t) - r_j (t)| < \delta} (r_j (t + \Delta t) - r_j (t)) + \Xi_i (t + \Delta t, t). \label{supp_eq_Vicsek_like_model_004_001}
\end{align}
However, it is time-consuming to solve Eq.~\eqref{supp_eq_Vicsek_like_model_004_001} since it involves a backward term.
Assuming $r_j (t + \Delta t) - r_j (t) = r_j (t) - r_j (t - \Delta t)$, we have
\begin{align}
  \alpha_i r_i (t + \Delta t) &= (\alpha_i - \beta_i \Delta t) r_i (t) + \frac{\gamma_i}{N_{i, \delta}} \sum_{j: |r_i (t) - r_j (t)| < \delta} (r_j (t) - r_j (t - \Delta t)) + \Xi_i (t + \Delta t, t).
\end{align}
Finally, we obtained Eq.~\eqref{main_eq_Vicsek_like_model_001_001}.

\subsection*{Availability of materials and data
}
Raw S\&P 500 stock price data between 2004 and 2020 are available at the Wharton Research Data Services (WRDS) \url{https://wrds-www.wharton.upenn.edu/}. The repository currently supports over $75000$ active researchers over $500$ institutions. The code bundle is available at \url{https://github.com/pholur/stock-market}.

\bibliography{citations}

\end{document}